\documentclass[letter,reprint]{revtex4-2}
\usepackage[utf8]{inputenc}
\usepackage[english]{babel}
\usepackage{geometry}
\usepackage{amsmath}
\usepackage{amssymb}
\usepackage{float}
\usepackage{graphicx,color}
\usepackage{hyperref}
\usepackage{natbib}

\begin{document}
\renewcommand{\figurename}{FIG.}

\title{Corrugation-dominated mechanical softening of defect-engineered graphene}
\author{Wael Joudi$^{1,2,*}$, Rika Saskia Windisch$^3$, Alberto Trentino$^{1,2}$, Diana Propst$^{1,2}$, Jacob Madsen$^{1}$,\\ Toma Susi$^{1}$, Clemens Mangler$^{1}$, Kimmo Mustonen$^{1}$, Florian Libisch$^3$ and Jani Kotakoski$^{1,*}$\\
$^1$University of Vienna, Faculty of Physics, Boltzmanngasse 5,\\ 1090 Vienna, Austria\\
$^2$University of Vienna, Vienna Doctoral School in Physics, Boltzmanngasse 5,\\ 1090 Vienna, Austria\\
$^3$Institute for Theoretical Physics, Faculty of Physics, Technical University of Vienna,\\ Wiedner Hauptstrasse 8-10, 1040 Vienna, Austria\\
$^*$Email: wael.joudi@univie.ac.at, jani.kotakoski@univie.ac.at}
\date{\today}

\begin{abstract}
We measure the two-dimensional elastic modulus $E^{\text{2D}}$ of atomically clean defect-engineered graphene with a known defect distribution and density in correlated ultra-high vacuum experiments.
The vacancies are introduced via low-energy ($<$ 200~eV) Ar ion irradiation and the atomic structure is obtained via semi-autonomous scanning transmission electron microscopy and image analysis.
Based on atomic force microscopy nanoindentation measurements, a decrease of $E^{\text{2D}}$ from $286$ to $158$~N/m is observed when measuring the same graphene membrane before and after an ion irradiation-induced vacancy density of $1.0 \times 10^{13}$~cm$^{-2}$. 
This decrease is significantly greater than what is predicted by most theoretical studies and in stark contrast to some measurements presented in the literature.
With the assistance of atomistic simulations, we show that this softening is mostly due to corrugations caused by local strain at vacancies with two or more missing atoms, while the influence of single vacancies is negligible.
We further demonstrate that the opposite effect can be measured when surface contamination is not removed before defect engineering.
\end{abstract} 

\maketitle

Experiments by Lee {\it et al.} revealed the exceptionally high intrinsic stiffness of monolayer graphene in 2008~\cite{Lee2008} based on atomic force microscopy (AFM) nanoindentation.
The 2D elastic modulus $E^{\text{2D}}$ was reported to be 340~N/m, which corresponds to a Young's modulus of 1~TPa assuming the inter-layer distance of graphite can be used as the thickness of graphene.
Several studies~\cite{Hao2011,Ito2012,Mortazavi2012,Tapia2012,Jing2012,Tserpes2012,Dettori2012,Mortazavi2013,Liang2015,Li2019,Zandiatashbar2014,Lopez-Polin2015,Chu2019,Kvashnin2015,Qin2016,Suk2020,Dai2019} have also explored the impact of lattice imperfections on the mechanical properties of graphene.
However, the results are not fully consistent.

Specifically, vacancy-type defects introduced via low energy Ar ions have been reported to increase $E^{\text{2D}}$ up to a maximum of 550~N/m at 0.2\% vacancy concentration (after which it decreases again)~\cite{Lopez-Polin2015}.
In contrast, low irradiation fluence with oxygen plasma was shown to not cause a clear change in $E^{\text{2D}}$, while at higher fluences $E^{\text{2D}}$ starts to decrease~\cite{Zandiatashbar2014}.
Moreover, for boron-doped graphene, an immediate decrease in $E^{\text{2D}}$ was reported after the transition from substitutional defects to vacancies~\cite{Dai2019}.
Similarly, simulations predict different outcomes for defect-engineered graphene.
Although most computational studies show a gradual decrease in $E^{\text{2D}}$ with increasing vacancy density~\cite{Tserpes2012,Dettori2012,Mortazavi2013,Liang2015,Li2019,Jing2012,Ito2012,Hao2011,Mortazavi2012,Tapia2012}, also an increase has been reported at low vacancy densities~\cite{Kvashnin2015,Kvashnin2010}. 
However, interestingly, corrugation results in a more drastic decrease of $E^{\text{2D}}$~\cite{Ruiz-Vargas2011,Qin2016,Suk2020,Wang2020}. 

In this study, we determine the relationship between the exact atomic structure of defect-engineered graphene and its 2D elastic modulus combining scanning transmission electron microscopy (STEM) medium-angle annular dark-field imaging (MAADF) and AFM nanoindentation.
We establish a direct correlation by performing the experiments in a vacuum system containing all instruments used~\cite{Mangler2022}.
Thus, AFM nanoindentation measurements are performed on atomically clean samples of known atomic structure.
These measurements reveal a significant decrease of $E^{\text{2D}}$ with increasing vacancy density. 
We present a model where the main contribution for the observed softening comes from vacancy-induced corrugation.
Molecular dynamics (MD) simulations using machine-learned force fields confirm this corrugation-dominated material softening, and suggest vacancies with two or more missing atoms as the dominant source of the increased corrugation.

\begin{figure}[b]
\centering
\includegraphics[width=0.49\textwidth]{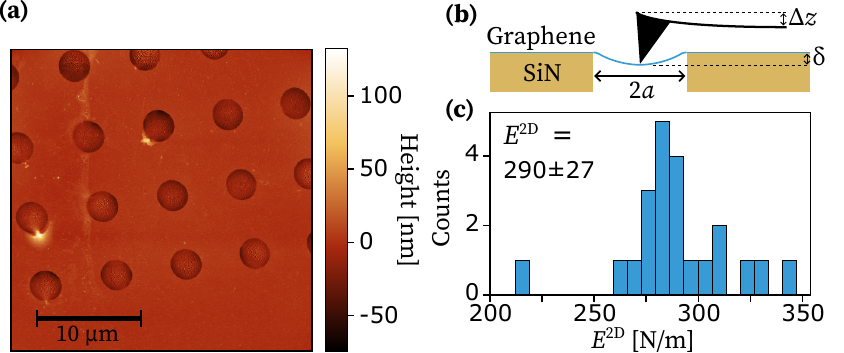}
\caption{\textbf{AFM nanoindentation of pristine graphene.} (a) AFM topography image of graphene supported by a perforated SiN membrane, (b) schematic illustration of the AFM nanoindentation measurement and (c) distribution of $E^{\text{2D}}$ measured on pristine graphene.}
\label{fig:AFM nanoindentation}
\end{figure}

The as-prepared graphene samples were inserted into the interconnected vacuum system~\cite{Mangler2022} through an overnight bake at 160$^\circ$C  (see End Matter for details about the methods).
An example AFM topography image is shown in Fig.~\ref{fig:AFM nanoindentation}(a).
AFM nanoindentation was performed on these graphene-covered holes (graphene drumheads), as shown schematically in Fig.~\ref{fig:AFM nanoindentation}(b).
The results for $E^{\text{2D}}$ of pristine graphene are summarized in Fig.~\ref{fig:AFM nanoindentation}(c). 
The Gaussian distribution has a mean value of $290\pm27$~N/m, which corresponds to a Young's modulus of $868\pm81$~GPa using the interlayer distance of graphite (3.34~{\AA}~\cite{Baskin1955}) as the thickness of graphene. 
The $E^{\text{2D}}$ values are in line with literature~\cite{Lee2008}, although the maximum of our statistical distribution is shifted toward lower values. 

\begin{figure*}[t]
\includegraphics[width=0.8\textwidth]{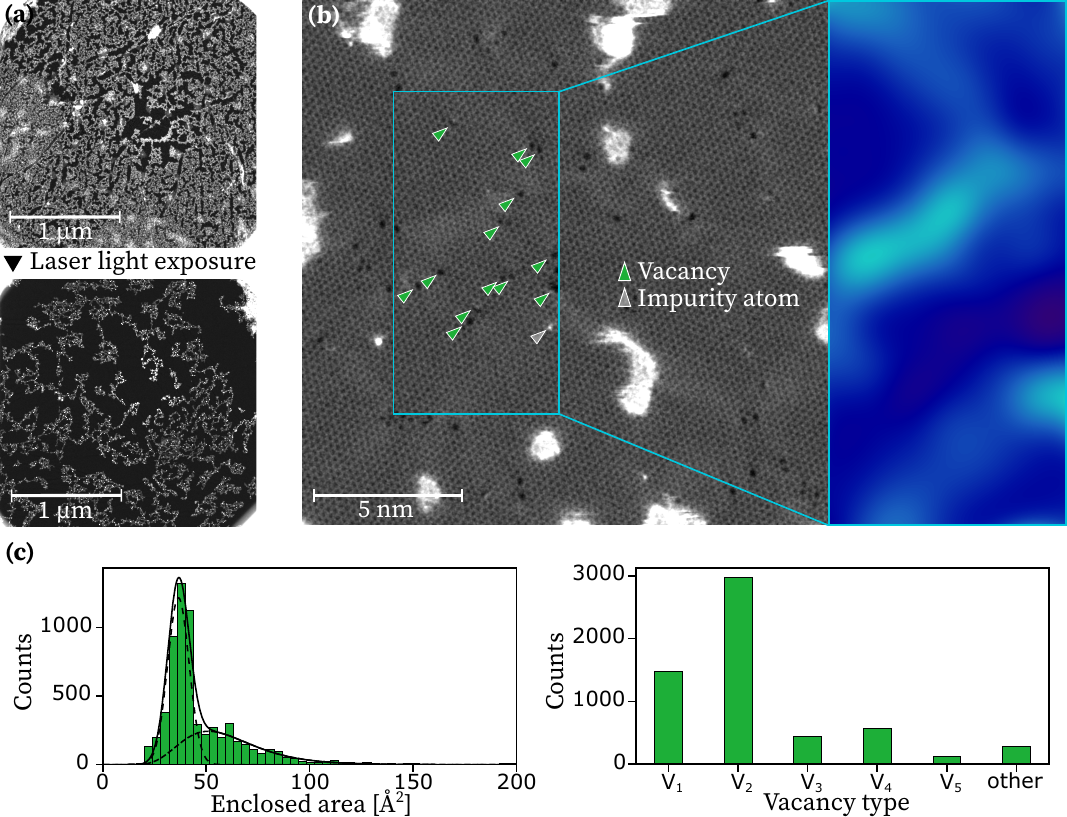}
\caption{\textbf{Defect-engineered graphene.}
(a) STEM-MAADF images of freestanding graphene before and after laser light exposure, (b) STEM-MAADF image of an area with vacancies at an areal density of $4.0 \times 10^{12}$~cm$^{-2}$ and some impurity atoms.
The magnification (with a Gaussian blur of $17$~px, the original image has $4096 \times 4096$~px of which a crop of ca. 1/4 is shown) highlights the morphological roughness.
(c) Vacancy area and type distributions across all samples ($V_i$ stands for a vacancy with $i$ missing atoms).}
\label{fig:STEM}
\end{figure*}

Surface contamination was removed using laser light illumination, resulting in most of the graphene lattice being exposed (Fig.~\ref{fig:STEM}(a)).
However, some contamination lines made up of spherical metallic nanoparticles remain visible as bright features in Fig.~\ref{fig:STEM}(a).
Next, Ar irradiation ($<200$~eV) was used~\cite{Alberto2021} to introduce vacancies of different types, as shown in Fig.~\ref{fig:STEM}(b).
Here, graphene is visible as the gray honeycomb lattice, which is disrupted by vacancies (seen due to enlarged carbon rings that appear as black spots).
Also a few impurity atoms with a negligible density appear embedded in the lattice (bright dots in $Z$-contrast images, where brightness scales with the atomic number~\cite{Krivanek2015}).
Some carbon-based contamination also remains, visible as the bright features of asymmetric shapes.
Finally, the brightness within the atomically clean graphene shows also subtle spatial variations due to alterations in the projected interatomic distance.
This arises from local corrugation, excemplified by the magnification of the area marked by the turquoise box with applied gaussian blur shown on the right-hand-side of Fig.~\ref{fig:STEM}(b).

The semi-automatically acquired images of defect-engineered graphene~\cite{Alberto2021,Mittelberger2018} were quantified in terms of their atomic structure by the CNN-based analyzer, as summarized in Fig.~\ref{fig:STEM}(c). 
Throughout all samples, a total of 5865 vacancies were recognized.
For vacancy densities up to $3.4 \times 10^{13}$~cm$^{-2}$, the areal size distributions consist of a Gaussian part and a log-normal part, which is consistent with our previous report~\cite{Alberto2021}.
With our irradiation parameters, the most abundant vacancy structures are single and double vacancies.
Defects that were only partially visible in the recorded images were neglected.
The individual statistical distributions are shown in the Supplemental Material~\cite{supplement}.

\begin{figure}[t]
\centering
\includegraphics[width=0.47\textwidth]{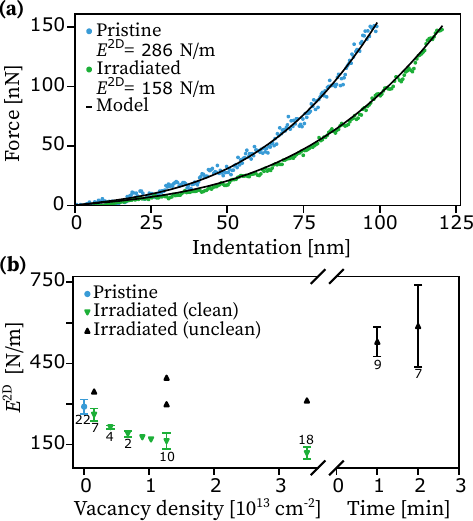}
\caption{\textbf{\boldmath{$E^{\text{2D}}$} of defect-engineered graphene.} (a) AFM nanoindentation results before and after ion irradiation and (b) $E^{\text{2D}}$ as a function of vacancy density or irradiation time. The error bars show the standard deviation and the numbers next to them represent the number of graphene drumheads measured for the corresponding data point.}
\label{fig: Nanoindentation_results}
\end{figure}

Contrary to previous measurements~\cite{Lopez-Polin2015}, the vacancies here cause a measurable decrease in the mechanical stiffness~\cite{Joudi2022}.
Fig.~\ref{fig: Nanoindentation_results}(a) shows a force-indentation curve of the same graphene drumhead before and after irradiation (vacancy density of $1.0 \times 10^{13}$~cm$^{-2}$),
revealing a decrease in $E^{\text{2D}}$ from $286$~N/m to $158$~N/m. 
We find that graphene drumheads cleaned prior to irradiation have a decreasing $E^{\text{2D}}$ with increasing vacancy density, while uncleaned or partially cleaned ones show a stiffness increase (Fig.~\ref{fig: Nanoindentation_results}(b)).
Note that for uncleaned samples the atomic structure is obstructed by contamination and therefore the $E^\text{2D}$ values are reported as a function of irradiation time rather than vacancy density.
Therefore it seems likely that the reported increase in $E^\text{2D}$~\cite{Lopez-Polin2015} is due to contamination build-up~\cite{Li2016,Li2013,Amadei2014,Kozbial20141,Kozbial20142,Ashraf2014,Wei2015,Muksch2015,Smith1998}.
However, the decrease in $E^{\text{2D}}$ seen in Fig.~\ref{fig: Nanoindentation_results}(b) is similar to what has been reported for oxygen plasma irradiation~\cite{Zandiatashbar2014}, which is additionally known to etch carbon-based surface contamination~\cite{Ponath2013,Baker1980,Chan2001,Oh2016,Isabell1999,Li1997}.
Therefore, the latter experiment should correspond to our experiments with clean graphene.

While also most theoretical studies report a moderate decrease in $E^{\text{2D}}$ with increasing vacancy density~\cite{Tserpes2012,Dettori2012,Mortazavi2013,Liang2015,Li2019,Jing2012,Ito2012,Hao2011,Mortazavi2012,Tapia2012}, an increase in $E^{\text{2D}}$ similar to the experimental findings of Ref.~\cite{Lopez-Polin2015} has also been reported, where the nanoindentation process was simulated on a $13.3$-nm graphene membrane~\cite{Kvashnin2015}. However, this effect disappeared for tensile loading of the membrane, which is more in line with the experimental situation, where the area with indenter-induced curvature is negligible compared to the size of the whole drumhead.

\begin{figure}[ht]
\centering
\includegraphics[width=0.47\textwidth]{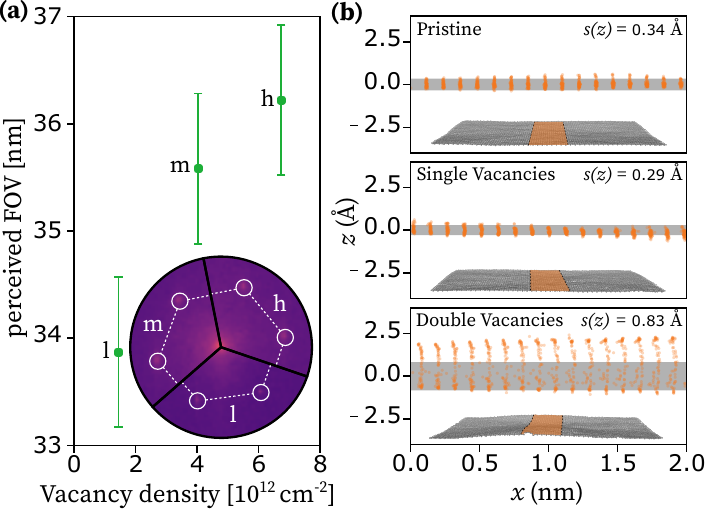}
\caption{\textbf{Vacancy-induced corrugation.}
(a) Perceived field of view (FOV) with increasing vacancy density.
The inset shows a composite image of the FFTs used for calculating the FOV at low (l), medium (m) and high (h) vacancy density.
The errorbars correspond to the estimated uncertainty.
(b) Cross-sectional views of simulated membranes at a vacancy concentration of 0.83\% (areal density of $3.2 \times 10^{13}$~cm$^{-2}$).
The orange dots correspond to carbon atoms while the grey overlay represents the corresponding  standard deviation $s(z)$, that was determined from the entire $12\times12$~nm$^2$ membrane (the cross-sectional views corresponds to a $2$~nm wide cutout).
The insets (created using VESTA~\cite{VESTA}) show the simulated graphene structures, where the orange area corresponds to the cutout (larger views are presented in the Supplemental Material~\cite{supplement}).}
\label{fig: corrugation_sim}
\end{figure}

In our study an unexpectedly steep and non-linear decrease in $E^{\text{2D}}$ is observed.
Interestingly, a similar steep decrease has been reported to be caused by corrugation of graphene due to small grain sizes or topological defects~\cite{Ruiz-Vargas2011,Qin2016,Suk2020,Wang2020}.
Vacancies have been similarly shown to corrugate graphene as a relaxation response to local strain introduced by bond rearrangement~\cite{Kotakoski2014,Thiemann2021}.
Since STEM-MAADF images (see Fig.~\ref{fig:STEM}(b)) suggest the presence of corrugation, we turn to this effect.
A more quantitative analysis is presented in Fig.\ref{fig: corrugation_sim}(a), where we estimate the size of the imaged area based on the fast Fourier transform (FFT) and the known lattice constant of graphene, for samples with three different defect densities (corresponding FFTs are shown in the inset). 
All images were recorded with a nominal scan size (field of view, FOV) of 32~nm. 
Comparing the FFTs reveals variations in the distance to the center, visualized by the white dashed lines not perfectly merging at the boundaries of the sectors, suggesting different lattice constants. 
However, since the physical lattice constant of graphene has not changed, these differences must arise from increasing corrugation that leads to shortening of the projected interatomic distances with increasing vacancy density~\cite{Lehtinen2013}.

For further insight, we turn to simulations of pristine and defective (single and double vacancies) graphene.
We consider membranes with periodic boundary conditions in the plane, with an active size of approximately $12 \times 12$~nm$^2$ (corresponding to ca. $6,000$ atoms) equilibrated at $400$~K (to accelerate structural relaxation), with the cell being relaxed in the in-plane directions. 
Fig.~\ref{fig: corrugation_sim}(b) shows example structures for each case from a cross-sectional view through a $2$~nm wide cutout.
A comparison reveals that while double vacancies substantially increases corrugation, single vacancies cause no clear change.
Repeating the simulation for $60$ different structures confirms this result.
We obtain standard deviations of the out-of-plane coordinate of the atoms ($s(z)$) of $0.36 \pm 0.08$~\AA, $0.35 \pm 0.08$~\AA~ and $0.64 \pm 0.17$~\AA~for pristine graphene, graphene with single vacancies and graphene with double vacancies, respectively.

This observation can be understood based on the Jahn-Teller distortion~\cite{PhysRevB.98.075439} that in single vacancies leads to the formation of one pentagon by bridging together two of the three carbon atoms with a dangling bond.
However, at a finite temperature, thermal activation allows this bond to switch between all three possible carbon-carbon pairs so that no direction is preferred.
In contrast, the most often found $V_2(585)$ double vacancy~\cite{Banhart2011} is associated with the formation of two such bonds that can only appear in one configuration.
This creates a permanent strain in the direction perpendicular to the vacancy axis, which further leads to local corrugation of the lattice.
For similar reasons we expect the same behavior for vacancies larger than double vacancies in terms of enhanced corrugation.

Based on this knowledge, we propose a semi-empirical model that incorporates both corrugation effects as well as changes due to the missing atoms into the estimation of $E^{\text{2D}}$ with increasing vacancy density
\begin{multline}
E^{\text{2D}}(n_{2+}) = E^{\text{2D}}_p \times \left(1 - A n_{2+} \right) 
\frac{1}{B \sqrt{n_{2+}} + 1},
\label{eq:model}
\end{multline}
where $A = \frac{\overline{N_m}}{C} \frac{1}{1.5 \rho_A}$ and $B = R_\varepsilon \alpha$. $E^{\text{2D}}_p$ is the 2D elastic modulus of pristine graphene of $290\pm27$~N/m, $n_{2+}$ is the density of vacancies with at least two missing atoms, $\rho_A = 38.2$~nm$^{-2}$ is the atomic areal density of graphene, the factor of $1.5$ comes from the fact that every carbon atom contributes $1.5$ $\sigma$-bonds, $\overline{N_m} = 4.405$ is the average number of missing $\sigma$-bonds per vacancy, $C = 0.804$ accounts for the contribution of single vacancies to the missing bonds, $\alpha = 1$~\AA~ is the corrugation amplitude~\cite{Kotakoski2014}, and $R_\varepsilon$ is the elastic energy ratio, which is a fitting parameter.
Here, $\overline{N_m}$ and $C$ are empirically determined constants based on the CNN analysis.
In equation (\ref{eq:model}) the linear factor accounts for the reduced $\sigma$-bond density due to missing carbon atoms, while the non-linear factor accounts for stiffness reduction due to corrugation caused by vacancies with two or more missing carbon atoms. 
This reduction scales with the amplitude of the corrugation as well as the areal density of the corrugation bumps. 
In the context of in-plane stiffness this corrugation density manifests as a one-dimensional spatial frequency of the corrugation bumps along the graphene sheet which is included as the average distance between the contributing defects, i.e., $\sqrt{n_{2+}}$. 
The corrugation amplitude as well as their spatial frequency are semi-empirically introduced as an inverse relationship where the only unknown parameter $R_\varepsilon$ accounts for softening of the locally corrugated area around each vacancy.

\begin{figure}[b]
\centering
\includegraphics[width=0.48\textwidth]{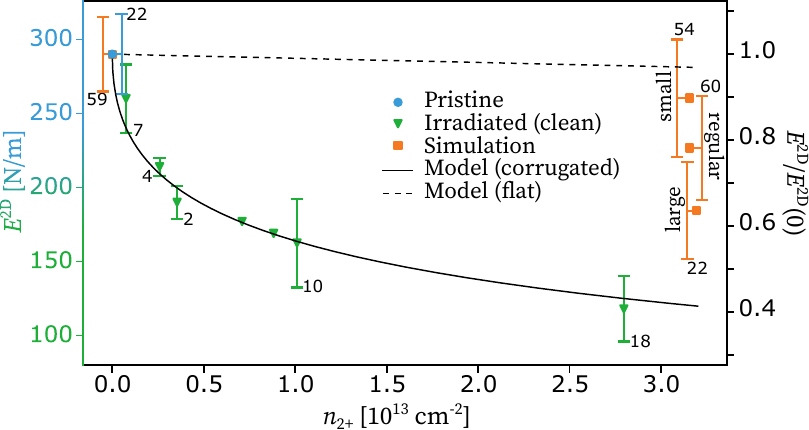}
\caption{\textbf{Softening due to corrugation.}
$E^{\text{2D}}$ as a function of the vacancy density for two or more missing atoms ($n_{2+}$).
The number next to each error bar represents the number of corresponding measurements.
The left $y$-axis represents the experimental values while the right $y$-axis represents the normalized values with respect to the pristine $E^{\text{2D}}$.
The data points are shown at the correct $x$-axis position, but in some cases the standard deviations are shifted.}
\label{fig: corrugation}
\end{figure}

The observed softening occurs due to stress-induced flattening of corrugations when applying a mechanical load.
While flattening the intrinsic corrugations of free-standing graphene~\cite{Meyer2007,Fasolino2007,Singh2022} doesn't require straining bonds, flattening vacancy-induced corrugation entails straining the bonds within the area of structural disorder, requiring deposition of elastic energy.
In other words, contrary to intrinsic corrugation, vacancy-induced corrugation can be seen as static.
The lower bond density and strength of $\sigma$-bonds associated with structurally disordered areas results in an effective softening of the material.
Fig.~\ref{fig: corrugation} shows the model displayed in Eq.~(\ref{eq:model}) applied to the data of clean irradiated graphene, where an $R_\varepsilon = 23.78$ provides a satisfactory fit.

The model supports our hypothesis that corrugation is the dominant factor softening graphene. 
This becomes clear when considering a flat membrane and attributing the reduction in stiffness solely to the missing $\sigma$-bonds, {\it i.e.}, when neglecting the corrugation factor (replacing the last factor in Eq.~(\ref{eq:model}) by 1), represented by the dashed line of Fig.~\ref{fig: corrugation}.
Similar to past theoretical studies~\cite{Tserpes2012,Dettori2012,Mortazavi2013,Liang2015,Li2019,Jing2012,Ito2012,Hao2011,Mortazavi2012,Tapia2012} this decrease is much more moderate than the experimental data.
Moreover, our model agrees with the findings obtained from simulations concerning the inconsequential impact of single vacancies on the corrugation.
The inclusion of single vacancies to the total vacancy density results in an overestimation of the corrugation effect leading to a visible deviation from the analytical model at lower densities (see Supplemental Material~\cite{supplement}), where according to the vacancy type distributions single vacancies have a relatively large abundance.
Lastly, the experimental data are compared with the simulation results (Fig.~\ref{fig: corrugation}), where similar to experimental findings, the simulations reveal a reduction in $E^{\text{2D}}$.
However, the observed reduction is not as large as in the experimental values.

We attribute the difference to finite-size effects due to the periodic boundary conditions used for our simulated membrane (regular, $158.5$~nm$^2$), which may lead to an underestimation of the long-range membrane corrugation ({\textit i.e.}, corrugations with a wavelength exceeding the simulation cell size cannot occur).
To demonstrate this, we perform simulations with both larger and smaller membrane sizes (56.6~nm$^2$ and 338.0~nm$^2$).
As expected, increasing the number of simulated atoms leads to a further decrease in the $E^{\text{2D}}$ values.
We note that the absolute values of the simulation results for $E^{\text{2D}}$ were significantly higher than the experimental values, and a bit higher than values predicted from DFT, and for this reason relative values are shown for the simulation results.
We attribute these differences to (i) the approximations in modeling, in particular assuming a perfect, entirely defect-free membrane as the reference value, and to (ii) the approximations within the machine-learned potential.
We do not expect any qualitative changes due to these approximations.

In conclusion, we have investigated the impact of vacancies in graphene on its two-dimensional elastic modulus. 
They are found to result in a decrease of $E^{\text{2D}}$, which is mainly due to corrugations associated with defects.
Atomistic simulations reveal that while single vacancies do not lead to enhanced corrugation, larger vacancy structures corrugate graphene and are therefore the main cause for the softening of defect-engineered graphene.
The results also reveal the importance of the removal of surface contamination in preparation for defect engineering, since the opposite effect on $E^{\text{2D}}$ is found when the contamination is not removed before irradiating the graphene.

\section*{Acknowledgments}

This research was funded in whole or in part by the Austrian Science Fund (FWF) projects [10.55776/P34797, 10.55776/COE5].

\section*{Data availability}
The data that support the findings of this article are openly available~\cite{phaidra}.

\appendix

\section*{End Matter}

\paragraph*{Sample preparation}
The {\it Easy Transfer} graphene from Graphenea Inc. graphene was put into de-ionized water floating on the water surface.
Next, the material was scooped out by custom-designed SiN TEM chips purchased from Silson Ltd followed by subsequent annealing in air at 150$^\circ$C.
Then, the Poly(methyl methacrylate) sacrificial layer was dissolved in an acetone bath at $50^\circ$C.
The necessity for a custom design stems from the need of a perforated support required for STEM that is also mechanically sufficiently rigid for the normal forces applied during AFM nanoindentation.
To the best of our knowledge, the latter is not fulfilled by most commercially available support chips due to lack of thickness of the SiN support membrane, as demonstrated in the Supplemental Material~\cite{supplement}, where the divergence between the curves acquired on the rigid Si wafer frame and the SiN membrane is a result of membrane bending.
Comparing the two shows drastically reduced membrane bending on the custom support with 1000~nm thickness and $90 \times 90$~$\mu$m$^2$ wide windows, which is sufficiently low for the forces applied during AFM nanoindentation.
Another advantage of the custom design is that the same sample location can be found in each instrument, which is enabled by a binary marker system of the SiN support windows arranged in a $3\times3$ array, where a missing hole represents a one and an existing hole represents a zero, as shown in the Supplemental Material~\cite{supplement}.

\paragraph*{Raman spectroscopy}
The Raman spectroscopy was performed on an alpha300 A developed by WITec GmbH using a $532$~nm laser light source at $5$~mW of power focused to a probe size of roughly $1$~$\mu$m in diameter, integrated over 0.5~s and accumulated 30 times.
The Supplemental Material~\cite{supplement} shows a light microscopy overview image of the sample, where each hole covered by free-standing graphene is classified based on its Raman spectrum, as highlighted in the magnification of the image.
Only pristine monolayers are used for this study, which are identified based on evaluation of the G peak ($1580$~cm$^{-1}$), the 2D peak ($\approx 2700$~cm$^{-1}$, dispersive) and the D peak ($\approx 1350$~cm$^{-1}$, dispersive)~\cite{Geim2006}.
An example of a graphene drumhead that fulfills these criteria is shown for hole 42 in the Supplemental Material~\cite{supplement}.

\paragraph*{Interconnected vacuum system}
The interconnected vacuum system contains a Nion UltraSTEM 100 with a base pressure of $10^{-10}$~mbar, an AFM device (AFSEM) by Quantum Design GmbH (low $10^{-9}$~mbar) and a target chamber (low $10^{-10}$~mbar) with a plasma source used for ion irradiation and a $6$~W continuous wave diode laser by Lasertack GmbH with $445$~nm wavelength used for contamination removal within UHV~\cite{Tripathi2017}.
Swift contamination deposition that would occur in ambient conditions~\cite{Li2016,Li2013,Amadei2014} is prevented by the uninterrupted UHV environment. 

\paragraph*{Sample cleaning}
The graphene surface was illuminated with laser light using powers at $17$\%, $27$\% or $42$\% of the maximum device power (depending on the sample).
Power adjustments were made in order to minimize the amount of mobile contamination and its electron beam-induced deposition~\cite{Dyck2017,Egerton2004,Catriona2012,Dyck2024} by overly excessive heating, that would prevent atomic resolution imaging. 

\paragraph*{Ar ion irradiation}
Ion irradiation was performed with a PCS-ECR-HO plasma source developed by SPECS Surface Nano Analysis GmbH.
Ar gas was leaked into the target chamber increasing the pressure to $10^{-6}$~mbar, followed by ionization and acceleration using a magnetron current of 16~mA.
The acceleration is purely caused by the sheath potential, resulting in kinetic energies of $<$ 200~eV.
The diameter of the divergent beam was estimated to be 9~cm at sample distance, thus a constant flux density was assumed throughout the 3~mm wide sample.

\paragraph*{Semi-autonomous STEM acquisition}
The MAADF detector covered a semi-angular range of 60-200~mrad and the convergence angle of the 60-keV electron beam was ca. 35~mrad.
For each sample, the selected areas in the order of $10^4$~nm$^2$ were defined by four corner points to establish the sample height at all positions through a bilinear interpolation.
The selected area was dissected into increments with a nominal size of $5 \times 5$~nm$^2$, which were semi-automatically imaged at atomic resolution.
In order to prevent overlap of the imaged areas, an offset of one or two images was used.
To minimize residual stage movement during image acquisition, a sleep time of two or three seconds between stage translation and start of the acquisition was implemented.

\paragraph*{Mechanical testing}
The $E^{\text{2D}}$ values were determined via AFM nanoindentation using an atomic force microscope.
The AFM device uses piezo-resistive cantilevers with Si or single crystal diamond tips.
The spring constants were determined by the cantilever geometry, frequency and $Q$-factor according to the method proposed by Sader \emph{et al.}~\cite{Sader1999}, and ranged from $50$~N/m up to $150$~N/m.
The maximum scan range of $30$~$\mu$m allows locating the individual graphene drumheads through dynamic mode imaging and the help of the sample map displayed in the Supplemental Material~\cite{supplement}.
The $E^{\text{2D}}$ was determined through a fit using
\begin{equation}
F(\delta) = (\sigma_0^{2D} \pi) \delta + \Bigg(\frac{E^{\text{2D}}}{a^2}\Bigg) \delta^3,
\label{eq:fit}
\end{equation}
where $F$ is the applied force $\sigma_0^{2D}$ is the pre-tension, $a$ is the radius of the drumhead and $\delta$ is the indentation depth of the graphene~\cite{Lopez-Polin2015,Lee2008}.
Here, $E^{\text{2D}}$ and $\sigma_0^{2D}$ are fit parameters.
The indentation depth is calculated as 
\begin{equation}
\delta = \left| z \right| - \Delta z,
\end{equation}
with $z$ being the height of the $z$-piezo position relative to its contact point position and $\Delta z$ being the vertical deflection of the cantilever.

\paragraph*{Molecular dynamics simulation}
A machine learning force field, trained via active learning for distorted and ruptured membranes, was developed for a total of $12,000$ MD steps over $12$ different membrane systems and a total of $585$ configurations evaluated by density functional theory (DFT) using the Perdew–Burke–Ernzerhof functional~\cite{Perdew1996}.
MD simulations were performed using VASP~\cite{VASP} with monoclinic simulation cells, and periodic boundary conditions in all directions were applied. To prevent interactions along the surface normal of the graphene sheet a vacuum of approximately $6$~nm was introduced. 
A total of $60$ simulations per vacancy density were performed.  
Each simulation started with a perfectly flat graphene sheet and a temperature of $T=400$ K (to accelerate the built-up of corrugation) in an isobaric-isothermal ensemble (NPT) using a Langevin thermostat.
The system was then equilibrated until the average corrugation no longer increased, which required $20$~ps in the case of pristine graphene and $40-60$~ps for defective graphene.
This was followed by $0.3$~ps of damped MD to relax the system, in order to obtain a good starting point for the subsequent procedure. Subsequently, the membrane was pulled at $T=300$~K by incrementally enlarging the in-plane cell size.
This included a total of $20$ strain steps, where in each step the in-plane cell size was increased by $0.05$\%.
Each step was simulated for $10$~ps, in order to ensure that the system has sufficient time to adapt to the increased volume.
The standard deviation of the atomic heights distribution was determined at each time step.
This distribution was sampled over all atoms and simulated systems.
The two-dimensional elastic modulus was then obtained by fitting the energy-strain curve with 
\begin{equation}
\frac{E(\epsilon_A)-E_0}{A_0}=\frac{E^{2D}}{4}\epsilon_A^2,
\end{equation}
where $\epsilon_A$ denotes the strain related to the area of the membrane, $A_0$ is the area of the membrane and $E_0$ corresponds to the energy of the relaxed membrane.

\bibliographystyle{apsrev4-1}
\bibliography{references}

\begin{thebibliography}{62}%
\makeatletter
\providecommand \@ifxundefined [1]{%
 \@ifx{#1\undefined}
}%
\providecommand \@ifnum [1]{%
 \ifnum #1\expandafter \@firstoftwo
 \else \expandafter \@secondoftwo
 \fi
}%
\providecommand \@ifx [1]{%
 \ifx #1\expandafter \@firstoftwo
 \else \expandafter \@secondoftwo
 \fi
}%
\providecommand \natexlab [1]{#1}%
\providecommand \enquote  [1]{``#1''}%
\providecommand \bibnamefont  [1]{#1}%
\providecommand \bibfnamefont [1]{#1}%
\providecommand \citenamefont [1]{#1}%
\providecommand \href@noop [0]{\@secondoftwo}%
\providecommand \href [0]{\begingroup \@sanitize@url \@href}%
\providecommand \@href[1]{\@@startlink{#1}\@@href}%
\providecommand \@@href[1]{\endgroup#1\@@endlink}%
\providecommand \@sanitize@url [0]{\catcode `\\12\catcode `\$12\catcode
  `\&12\catcode `\#12\catcode `\^12\catcode `\_12\catcode `\%12\relax}%
\providecommand \@@startlink[1]{}%
\providecommand \@@endlink[0]{}%
\providecommand \url  [0]{\begingroup\@sanitize@url \@url }%
\providecommand \@url [1]{\endgroup\@href {#1}{\urlprefix }}%
\providecommand \urlprefix  [0]{URL }%
\providecommand \Eprint [0]{\href }%
\providecommand \doibase [0]{http://dx.doi.org/}%
\providecommand \selectlanguage [0]{\@gobble}%
\providecommand \bibinfo  [0]{\@secondoftwo}%
\providecommand \bibfield  [0]{\@secondoftwo}%
\providecommand \translation [1]{[#1]}%
\providecommand \BibitemOpen [0]{}%
\providecommand \bibitemStop [0]{}%
\providecommand \bibitemNoStop [0]{.\EOS\space}%
\providecommand \EOS [0]{\spacefactor3000\relax}%
\providecommand \BibitemShut  [1]{\csname bibitem#1\endcsname}%
\let\auto@bib@innerbib\@empty
\bibitem [{\citenamefont {Lee}\ \emph {et~al.}(2008)\citenamefont {Lee},
  \citenamefont {Wei}, \citenamefont {Kysar},\ and\ \citenamefont
  {Hone}}]{Lee2008}%
  \BibitemOpen
  \bibfield  {author} {\bibinfo {author} {\bibfnamefont {C.}~\bibnamefont
  {Lee}}, \bibinfo {author} {\bibfnamefont {X.}~\bibnamefont {Wei}}, \bibinfo
  {author} {\bibfnamefont {J.~W.}\ \bibnamefont {Kysar}}, \ and\ \bibinfo
  {author} {\bibfnamefont {J.}~\bibnamefont {Hone}},\ }\href {\doibase
  10.1126/science.1157996} {\bibfield  {journal} {\bibinfo  {journal}
  {Science}\ }\textbf {\bibinfo {volume} {321}},\ \bibinfo {pages} {385}
  (\bibinfo {year} {2008})}\BibitemShut {NoStop}%
\bibitem [{\citenamefont {Hao}\ \emph {et~al.}(2011)\citenamefont {Hao},
  \citenamefont {Fang},\ and\ \citenamefont {Xu}}]{Hao2011}%
  \BibitemOpen
  \bibfield  {author} {\bibinfo {author} {\bibfnamefont {F.}~\bibnamefont
  {Hao}}, \bibinfo {author} {\bibfnamefont {D.}~\bibnamefont {Fang}}, \ and\
  \bibinfo {author} {\bibfnamefont {Z.}~\bibnamefont {Xu}},\ }\href {\doibase
  10.1063/1.3615290} {\bibfield  {journal} {\bibinfo  {journal} {Applied
  Physics Letters}\ }\textbf {\bibinfo {volume} {99}},\ \bibinfo {pages}
  {041901} (\bibinfo {year} {2011})}\BibitemShut {NoStop}%
\bibitem [{\citenamefont {Ito}\ and\ \citenamefont {Okamoto}(2012)}]{Ito2012}%
  \BibitemOpen
  \bibfield  {author} {\bibinfo {author} {\bibfnamefont {A.}~\bibnamefont
  {Ito}}\ and\ \bibinfo {author} {\bibfnamefont {S.}~\bibnamefont {Okamoto}},\
  }\href@noop {} {\bibfield  {journal} {\bibinfo  {journal} {Engineering
  Letters}\ }\textbf {\bibinfo {volume} {20}} (\bibinfo {year}
  {2012})}\BibitemShut {NoStop}%
\bibitem [{\citenamefont {Mortazavi}\ \emph {et~al.}(2012)\citenamefont
  {Mortazavi}, \citenamefont {Ahzi}, \citenamefont {Toniazzo},\ and\
  \citenamefont {Rémond}}]{Mortazavi2012}%
  \BibitemOpen
  \bibfield  {author} {\bibinfo {author} {\bibfnamefont {B.}~\bibnamefont
  {Mortazavi}}, \bibinfo {author} {\bibfnamefont {S.}~\bibnamefont {Ahzi}},
  \bibinfo {author} {\bibfnamefont {V.}~\bibnamefont {Toniazzo}}, \ and\
  \bibinfo {author} {\bibfnamefont {Y.}~\bibnamefont {Rémond}},\ }\href
  {\doibase https://doi.org/10.1016/j.physleta.2011.11.034} {\bibfield
  {journal} {\bibinfo  {journal} {Physics Letters A}\ }\textbf {\bibinfo
  {volume} {376}},\ \bibinfo {pages} {1146} (\bibinfo {year}
  {2012})}\BibitemShut {NoStop}%
\bibitem [{\citenamefont {Tapia}\ \emph {et~al.}(2012)\citenamefont {Tapia},
  \citenamefont {Peón-Escalante}, \citenamefont {Villanueva},\ and\
  \citenamefont {Avilés}}]{Tapia2012}%
  \BibitemOpen
  \bibfield  {author} {\bibinfo {author} {\bibfnamefont {A.}~\bibnamefont
  {Tapia}}, \bibinfo {author} {\bibfnamefont {R.}~\bibnamefont
  {Peón-Escalante}}, \bibinfo {author} {\bibfnamefont {C.}~\bibnamefont
  {Villanueva}}, \ and\ \bibinfo {author} {\bibfnamefont {F.}~\bibnamefont
  {Avilés}},\ }\href {\doibase
  https://doi.org/10.1016/j.commatsci.2011.12.013} {\bibfield  {journal}
  {\bibinfo  {journal} {Computational Materials Science}\ }\textbf {\bibinfo
  {volume} {55}},\ \bibinfo {pages} {255} (\bibinfo {year} {2012})}\BibitemShut
  {NoStop}%
\bibitem [{\citenamefont {Jing}\ \emph {et~al.}(2012)\citenamefont {Jing},
  \citenamefont {Xue}, \citenamefont {Ling}, \citenamefont {Shan},
  \citenamefont {Zhang}, \citenamefont {Zhou},\ and\ \citenamefont
  {Jiao}}]{Jing2012}%
  \BibitemOpen
  \bibfield  {author} {\bibinfo {author} {\bibfnamefont {N.}~\bibnamefont
  {Jing}}, \bibinfo {author} {\bibfnamefont {Q.}~\bibnamefont {Xue}}, \bibinfo
  {author} {\bibfnamefont {C.}~\bibnamefont {Ling}}, \bibinfo {author}
  {\bibfnamefont {M.}~\bibnamefont {Shan}}, \bibinfo {author} {\bibfnamefont
  {T.}~\bibnamefont {Zhang}}, \bibinfo {author} {\bibfnamefont
  {X.}~\bibnamefont {Zhou}}, \ and\ \bibinfo {author} {\bibfnamefont
  {Z.}~\bibnamefont {Jiao}},\ }\href {\doibase 10.1039/C2RA21228E} {\bibfield
  {journal} {\bibinfo  {journal} {RSC Adv.}\ }\textbf {\bibinfo {volume} {2}},\
  \bibinfo {pages} {9124} (\bibinfo {year} {2012})}\BibitemShut {NoStop}%
\bibitem [{\citenamefont {Tserpes}(2012)}]{Tserpes2012}%
  \BibitemOpen
  \bibfield  {author} {\bibinfo {author} {\bibfnamefont {K.}~\bibnamefont
  {Tserpes}},\ }\href {\doibase 10.1007/s00707-011-0594-8} {\bibfield
  {journal} {\bibinfo  {journal} {Acta Mechanica}\ }\textbf {\bibinfo {volume}
  {223}},\ \bibinfo {pages} {669–678} (\bibinfo {year} {2012})}\BibitemShut
  {NoStop}%
\bibitem [{\citenamefont {Dettori}\ \emph {et~al.}(2012)\citenamefont
  {Dettori}, \citenamefont {Cadelano},\ and\ \citenamefont
  {Colombo}}]{Dettori2012}%
  \BibitemOpen
  \bibfield  {author} {\bibinfo {author} {\bibfnamefont {R.}~\bibnamefont
  {Dettori}}, \bibinfo {author} {\bibfnamefont {E.}~\bibnamefont {Cadelano}}, \
  and\ \bibinfo {author} {\bibfnamefont {L.}~\bibnamefont {Colombo}},\ }\href
  {\doibase 10.1088/0953-8984/24/10/104020} {\bibfield  {journal} {\bibinfo
  {journal} {Journal of Physics: Condensed Matter}\ }\textbf {\bibinfo {volume}
  {24}},\ \bibinfo {pages} {104020} (\bibinfo {year} {2012})}\BibitemShut
  {NoStop}%
\bibitem [{\citenamefont {Mortazavi}\ and\ \citenamefont
  {Ahzi}(2013)}]{Mortazavi2013}%
  \BibitemOpen
  \bibfield  {author} {\bibinfo {author} {\bibfnamefont {B.}~\bibnamefont
  {Mortazavi}}\ and\ \bibinfo {author} {\bibfnamefont {S.}~\bibnamefont
  {Ahzi}},\ }\href {\doibase https://doi.org/10.1016/j.carbon.2013.07.017}
  {\bibfield  {journal} {\bibinfo  {journal} {Carbon}\ }\textbf {\bibinfo
  {volume} {63}},\ \bibinfo {pages} {460} (\bibinfo {year} {2013})}\BibitemShut
  {NoStop}%
\bibitem [{\citenamefont {Yingjing~Liang}\ and\ \citenamefont
  {Huan}(2015)}]{Liang2015}%
  \BibitemOpen
  \bibfield  {author} {\bibinfo {author} {\bibfnamefont {Q.~H.}\ \bibnamefont
  {Yingjing~Liang}}\ and\ \bibinfo {author} {\bibfnamefont {S.}~\bibnamefont
  {Huan}},\ }\href {\doibase 10.1080/01495739.2015.1040317} {\bibfield
  {journal} {\bibinfo  {journal} {Journal of Thermal Stresses}\ }\textbf
  {\bibinfo {volume} {38}},\ \bibinfo {pages} {926} (\bibinfo {year}
  {2015})}\BibitemShut {NoStop}%
\bibitem [{\citenamefont {Li}\ \emph {et~al.}(2019)\citenamefont {Li},
  \citenamefont {Deng}, \citenamefont {Zheng}, \citenamefont {Zhang},
  \citenamefont {Liao},\ and\ \citenamefont {Zhou}}]{Li2019}%
  \BibitemOpen
  \bibfield  {author} {\bibinfo {author} {\bibfnamefont {M.}~\bibnamefont
  {Li}}, \bibinfo {author} {\bibfnamefont {T.}~\bibnamefont {Deng}}, \bibinfo
  {author} {\bibfnamefont {B.}~\bibnamefont {Zheng}}, \bibinfo {author}
  {\bibfnamefont {Y.}~\bibnamefont {Zhang}}, \bibinfo {author} {\bibfnamefont
  {Y.}~\bibnamefont {Liao}}, \ and\ \bibinfo {author} {\bibfnamefont
  {H.}~\bibnamefont {Zhou}},\ }\href {\doibase 10.3390/nano9030347} {\bibfield
  {journal} {\bibinfo  {journal} {Nanomaterials}\ }\textbf {\bibinfo {volume}
  {9}},\ \bibinfo {pages} {347} (\bibinfo {year} {2019})}\BibitemShut {NoStop}%
\bibitem [{\citenamefont {Zandiatashbar}\ \emph {et~al.}(2014)\citenamefont
  {Zandiatashbar}, \citenamefont {Lee}, \citenamefont {An}, \citenamefont
  {Lee}, \citenamefont {Mathew}, \citenamefont {Terrones}, \citenamefont
  {Hayashi}, \citenamefont {Picu}, \citenamefont {Hone},\ and\ \citenamefont
  {Koratkar}}]{Zandiatashbar2014}%
  \BibitemOpen
  \bibfield  {author} {\bibinfo {author} {\bibfnamefont {A.}~\bibnamefont
  {Zandiatashbar}}, \bibinfo {author} {\bibfnamefont {G.~H.}\ \bibnamefont
  {Lee}}, \bibinfo {author} {\bibfnamefont {S.~J.}\ \bibnamefont {An}},
  \bibinfo {author} {\bibfnamefont {S.}~\bibnamefont {Lee}}, \bibinfo {author}
  {\bibfnamefont {N.}~\bibnamefont {Mathew}}, \bibinfo {author} {\bibfnamefont
  {M.}~\bibnamefont {Terrones}}, \bibinfo {author} {\bibfnamefont
  {T.}~\bibnamefont {Hayashi}}, \bibinfo {author} {\bibfnamefont {C.~R.}\
  \bibnamefont {Picu}}, \bibinfo {author} {\bibfnamefont {J.}~\bibnamefont
  {Hone}}, \ and\ \bibinfo {author} {\bibfnamefont {N.}~\bibnamefont
  {Koratkar}},\ }\href {\doibase 10.1038/ncomms4186} {\bibfield  {journal}
  {\bibinfo  {journal} {Nature Communications}\ }\textbf {\bibinfo {volume}
  {5}},\ \bibinfo {pages} {3186} (\bibinfo {year} {2014})}\BibitemShut
  {NoStop}%
\bibitem [{\citenamefont {L{\'{o}}pez-Pol{\'{i}}n}\ \emph
  {et~al.}(2015)\citenamefont {L{\'{o}}pez-Pol{\'{i}}n}, \citenamefont
  {G{\'{o}}mez-Navarro}, \citenamefont {Parente}, \citenamefont {Guinea},
  \citenamefont {Katsnelson}, \citenamefont {P{\'{e}}rez-Murano},\ and\
  \citenamefont {G{\'{o}}mez-Herrero}}]{Lopez-Polin2015}%
  \BibitemOpen
  \bibfield  {author} {\bibinfo {author} {\bibfnamefont {G.}~\bibnamefont
  {L{\'{o}}pez-Pol{\'{i}}n}}, \bibinfo {author} {\bibfnamefont
  {C.}~\bibnamefont {G{\'{o}}mez-Navarro}}, \bibinfo {author} {\bibfnamefont
  {V.}~\bibnamefont {Parente}}, \bibinfo {author} {\bibfnamefont
  {F.}~\bibnamefont {Guinea}}, \bibinfo {author} {\bibfnamefont {M.~I.}\
  \bibnamefont {Katsnelson}}, \bibinfo {author} {\bibfnamefont
  {F.}~\bibnamefont {P{\'{e}}rez-Murano}}, \ and\ \bibinfo {author}
  {\bibfnamefont {J.}~\bibnamefont {G{\'{o}}mez-Herrero}},\ }\href {\doibase
  10.1038/nphys3183} {\bibfield  {journal} {\bibinfo  {journal} {Nature
  Physics}\ }\textbf {\bibinfo {volume} {11}},\ \bibinfo {pages} {26} (\bibinfo
  {year} {2015})}\BibitemShut {NoStop}%
\bibitem [{\citenamefont {Chu}\ \emph {et~al.}(2019)\citenamefont {Chu},
  \citenamefont {Shi},\ and\ \citenamefont {Braun}}]{Chu2019}%
  \BibitemOpen
  \bibfield  {author} {\bibinfo {author} {\bibfnamefont {L.}~\bibnamefont
  {Chu}}, \bibinfo {author} {\bibfnamefont {J.}~\bibnamefont {Shi}}, \ and\
  \bibinfo {author} {\bibfnamefont {R.}~\bibnamefont {Braun}},\ }\href
  {\doibase 10.1016/j.physe.2019.01.023} {\bibfield  {journal} {\bibinfo
  {journal} {Physica E: Low-dimensional Systems and Nanostructures}\ }\textbf
  {\bibinfo {volume} {110}},\ \bibinfo {pages} {115} (\bibinfo {year}
  {2019})}\BibitemShut {NoStop}%
\bibitem [{\citenamefont {Kvashnin}\ and\ \citenamefont
  {Sorokin}(2015)}]{Kvashnin2015}%
  \BibitemOpen
  \bibfield  {author} {\bibinfo {author} {\bibfnamefont {D.}~\bibnamefont
  {Kvashnin}}\ and\ \bibinfo {author} {\bibfnamefont {P.}~\bibnamefont
  {Sorokin}},\ }\href {\doibase 10.1021/acs.jpclett.5b00740} {\bibfield
  {journal} {\bibinfo  {journal} {The Journal of Physical Chemistry Letters}\
  }\textbf {\bibinfo {volume} {6}},\ \bibinfo {pages} {2384} (\bibinfo {year}
  {2015})}\BibitemShut {NoStop}%
\bibitem [{\citenamefont {Qin}\ \emph {et~al.}(2016)\citenamefont {Qin},
  \citenamefont {Sun}, \citenamefont {Liu},\ and\ \citenamefont
  {Liu}}]{Qin2016}%
  \BibitemOpen
  \bibfield  {author} {\bibinfo {author} {\bibfnamefont {H.}~\bibnamefont
  {Qin}}, \bibinfo {author} {\bibfnamefont {Y.}~\bibnamefont {Sun}}, \bibinfo
  {author} {\bibfnamefont {J.~Z.}\ \bibnamefont {Liu}}, \ and\ \bibinfo
  {author} {\bibfnamefont {Y.}~\bibnamefont {Liu}},\ }\href {\doibase
  https://doi.org/10.1016/j.carbon.2016.07.014} {\bibfield  {journal} {\bibinfo
   {journal} {Carbon}\ }\textbf {\bibinfo {volume} {108}},\ \bibinfo {pages}
  {204} (\bibinfo {year} {2016})}\BibitemShut {NoStop}%
\bibitem [{\citenamefont {Suk}\ \emph {et~al.}(2020)\citenamefont {Suk},
  \citenamefont {Hao}, \citenamefont {Liechti},\ and\ \citenamefont
  {Ruoff}}]{Suk2020}%
  \BibitemOpen
  \bibfield  {author} {\bibinfo {author} {\bibfnamefont {J.~W.}\ \bibnamefont
  {Suk}}, \bibinfo {author} {\bibfnamefont {Y.}~\bibnamefont {Hao}}, \bibinfo
  {author} {\bibfnamefont {K.~M.}\ \bibnamefont {Liechti}}, \ and\ \bibinfo
  {author} {\bibfnamefont {R.~S.}\ \bibnamefont {Ruoff}},\ }\href {\doibase
  10.1021/acs.chemmater.0c01660} {\bibfield  {journal} {\bibinfo  {journal}
  {Chemistry of Materials}\ }\textbf {\bibinfo {volume} {32}},\ \bibinfo
  {pages} {6078} (\bibinfo {year} {2020})}\BibitemShut {NoStop}%
\bibitem [{\citenamefont {Dai}\ \emph {et~al.}(2019)\citenamefont {Dai},
  \citenamefont {Wang}, \citenamefont {Zheng}, \citenamefont {Wang},
  \citenamefont {Zhang}, \citenamefont {Qi}, \citenamefont {Tan}, \citenamefont
  {Liu}, \citenamefont {Xu}, \citenamefont {Li}, \citenamefont {Cheng},\ and\
  \citenamefont {Zhang}}]{Dai2019}%
  \BibitemOpen
  \bibfield  {author} {\bibinfo {author} {\bibfnamefont {Z.}~\bibnamefont
  {Dai}}, \bibinfo {author} {\bibfnamefont {G.}~\bibnamefont {Wang}}, \bibinfo
  {author} {\bibfnamefont {Z.}~\bibnamefont {Zheng}}, \bibinfo {author}
  {\bibfnamefont {Y.}~\bibnamefont {Wang}}, \bibinfo {author} {\bibfnamefont
  {S.}~\bibnamefont {Zhang}}, \bibinfo {author} {\bibfnamefont
  {X.}~\bibnamefont {Qi}}, \bibinfo {author} {\bibfnamefont {P.}~\bibnamefont
  {Tan}}, \bibinfo {author} {\bibfnamefont {L.}~\bibnamefont {Liu}}, \bibinfo
  {author} {\bibfnamefont {Z.}~\bibnamefont {Xu}}, \bibinfo {author}
  {\bibfnamefont {Q.}~\bibnamefont {Li}}, \bibinfo {author} {\bibfnamefont
  {Z.}~\bibnamefont {Cheng}}, \ and\ \bibinfo {author} {\bibfnamefont
  {Z.}~\bibnamefont {Zhang}},\ }\href {\doibase
  https://doi.org/10.1016/j.carbon.2019.03.014} {\bibfield  {journal} {\bibinfo
   {journal} {Carbon}\ }\textbf {\bibinfo {volume} {147}},\ \bibinfo {pages}
  {594} (\bibinfo {year} {2019})}\BibitemShut {NoStop}%
\bibitem [{\citenamefont {Kvashnin}\ \emph {et~al.}(2010)\citenamefont
  {Kvashnin}, \citenamefont {Sorokin},\ and\ \citenamefont
  {Kvashnin}}]{Kvashnin2010}%
  \BibitemOpen
  \bibfield  {author} {\bibinfo {author} {\bibfnamefont {A.~G.}\ \bibnamefont
  {Kvashnin}}, \bibinfo {author} {\bibfnamefont {P.~B.}\ \bibnamefont
  {Sorokin}}, \ and\ \bibinfo {author} {\bibfnamefont {D.~G.}\ \bibnamefont
  {Kvashnin}},\ }\href {\doibase 10.1080/1536383X.2010.488160} {\bibfield
  {journal} {\bibinfo  {journal} {Fullerenes Nanotubes and Carbon
  Nanostructures}\ }\textbf {\bibinfo {volume} {18}},\ \bibinfo {pages} {497}
  (\bibinfo {year} {2010})}\BibitemShut {NoStop}%
\bibitem [{\citenamefont {Ruiz-Vargas}\ \emph {et~al.}(2011)\citenamefont
  {Ruiz-Vargas}, \citenamefont {Zhuang}, \citenamefont {Huang}, \citenamefont
  {{Van Der Zande}}, \citenamefont {Garg}, \citenamefont {McEuen},
  \citenamefont {Muller}, \citenamefont {Hennig},\ and\ \citenamefont
  {Park}}]{Ruiz-Vargas2011}%
  \BibitemOpen
  \bibfield  {author} {\bibinfo {author} {\bibfnamefont {C.~S.}\ \bibnamefont
  {Ruiz-Vargas}}, \bibinfo {author} {\bibfnamefont {H.~L.}\ \bibnamefont
  {Zhuang}}, \bibinfo {author} {\bibfnamefont {P.~Y.}\ \bibnamefont {Huang}},
  \bibinfo {author} {\bibfnamefont {A.~M.}\ \bibnamefont {{Van Der Zande}}},
  \bibinfo {author} {\bibfnamefont {S.}~\bibnamefont {Garg}}, \bibinfo {author}
  {\bibfnamefont {P.~L.}\ \bibnamefont {McEuen}}, \bibinfo {author}
  {\bibfnamefont {D.~A.}\ \bibnamefont {Muller}}, \bibinfo {author}
  {\bibfnamefont {R.~G.}\ \bibnamefont {Hennig}}, \ and\ \bibinfo {author}
  {\bibfnamefont {J.}~\bibnamefont {Park}},\ }\href {\doibase
  10.1021/NL200429F} {\bibfield  {journal} {\bibinfo  {journal} {Nano Letters}\
  }\textbf {\bibinfo {volume} {11}},\ \bibinfo {pages} {2259} (\bibinfo {year}
  {2011})}\BibitemShut {NoStop}%
\bibitem [{\citenamefont {Wang}\ \emph {et~al.}(2020)\citenamefont {Wang},
  \citenamefont {Pang}, \citenamefont {Li},\ and\ \citenamefont
  {Lai}}]{Wang2020}%
  \BibitemOpen
  \bibfield  {author} {\bibinfo {author} {\bibfnamefont {R.}~\bibnamefont
  {Wang}}, \bibinfo {author} {\bibfnamefont {H.}~\bibnamefont {Pang}}, \bibinfo
  {author} {\bibfnamefont {M.}~\bibnamefont {Li}}, \ and\ \bibinfo {author}
  {\bibfnamefont {L.}~\bibnamefont {Lai}},\ }\href {\doibase
  10.3390/ma13051127} {\bibfield  {journal} {\bibinfo  {journal} {Materials}\
  }\textbf {\bibinfo {volume} {13}},\ \bibinfo {pages} {1127} (\bibinfo {year}
  {2020})}\BibitemShut {NoStop}%
\bibitem [{\citenamefont {Mangler}\ \emph {et~al.}(2022)\citenamefont
  {Mangler}, \citenamefont {Meyer}, \citenamefont {Mittelberger}, \citenamefont
  {Mustonen}, \citenamefont {Susi},\ and\ \citenamefont
  {Kotakoski}}]{Mangler2022}%
  \BibitemOpen
  \bibfield  {author} {\bibinfo {author} {\bibfnamefont {C.}~\bibnamefont
  {Mangler}}, \bibinfo {author} {\bibfnamefont {J.}~\bibnamefont {Meyer}},
  \bibinfo {author} {\bibfnamefont {A.}~\bibnamefont {Mittelberger}}, \bibinfo
  {author} {\bibfnamefont {K.}~\bibnamefont {Mustonen}}, \bibinfo {author}
  {\bibfnamefont {T.}~\bibnamefont {Susi}}, \ and\ \bibinfo {author}
  {\bibfnamefont {J.}~\bibnamefont {Kotakoski}},\ }\href {\doibase
  10.1017/S1431927622011023} {\bibfield  {journal} {\bibinfo  {journal}
  {Microscopy and Microanalysis}\ }\textbf {\bibinfo {volume} {28 (S1)}},\
  \bibinfo {pages} {2940} (\bibinfo {year} {2022})}\BibitemShut {NoStop}%
\bibitem [{\citenamefont {Baskin}\ and\ \citenamefont
  {Meyer}(1955)}]{Baskin1955}%
  \BibitemOpen
  \bibfield  {author} {\bibinfo {author} {\bibfnamefont {Y.}~\bibnamefont
  {Baskin}}\ and\ \bibinfo {author} {\bibfnamefont {L.}~\bibnamefont {Meyer}},\
  }\href {\doibase 10.1103/PhysRev.100.544} {\bibfield  {journal} {\bibinfo
  {journal} {Phys. Rev.}\ }\textbf {\bibinfo {volume} {100}},\ \bibinfo {pages}
  {544} (\bibinfo {year} {1955})}\BibitemShut {NoStop}%
\bibitem [{\citenamefont {Trentino}\ \emph {et~al.}(2021)\citenamefont
  {Trentino}, \citenamefont {Madsen}, \citenamefont {Mittelberger},
  \citenamefont {Mangler}, \citenamefont {Susi}, \citenamefont {Mustonen},\
  and\ \citenamefont {Kotakoski}}]{Alberto2021}%
  \BibitemOpen
  \bibfield  {author} {\bibinfo {author} {\bibfnamefont {A.}~\bibnamefont
  {Trentino}}, \bibinfo {author} {\bibfnamefont {J.}~\bibnamefont {Madsen}},
  \bibinfo {author} {\bibfnamefont {A.}~\bibnamefont {Mittelberger}}, \bibinfo
  {author} {\bibfnamefont {C.}~\bibnamefont {Mangler}}, \bibinfo {author}
  {\bibfnamefont {T.}~\bibnamefont {Susi}}, \bibinfo {author} {\bibfnamefont
  {K.}~\bibnamefont {Mustonen}}, \ and\ \bibinfo {author} {\bibfnamefont
  {J.}~\bibnamefont {Kotakoski}},\ }\href {\doibase
  10.1021/acs.nanolett.1c01214} {\bibfield  {journal} {\bibinfo  {journal}
  {Nano Letters}\ }\textbf {\bibinfo {volume} {21}},\ \bibinfo {pages} {5179}
  (\bibinfo {year} {2021})}\BibitemShut {NoStop}%
\bibitem [{\citenamefont {Krivanek}\ \emph {et~al.}(2015)\citenamefont
  {Krivanek}, \citenamefont {Lovejoy},\ and\ \citenamefont
  {Dellby}}]{Krivanek2015}%
  \BibitemOpen
  \bibfield  {author} {\bibinfo {author} {\bibfnamefont {O.}~\bibnamefont
  {Krivanek}}, \bibinfo {author} {\bibfnamefont {T.}~\bibnamefont {Lovejoy}}, \
  and\ \bibinfo {author} {\bibfnamefont {N.}~\bibnamefont {Dellby}},\ }\href
  {\doibase 10.1111/jmi.12254} {\bibfield  {journal} {\bibinfo  {journal}
  {Journal of microscopy}\ }\textbf {\bibinfo {volume} {259}},\ \bibinfo
  {pages} {165} (\bibinfo {year} {2015})}\BibitemShut {NoStop}%
\bibitem [{\citenamefont {Mittelberger}(2018)}]{Mittelberger2018}%
  \BibitemOpen
  \bibfield  {author} {\bibinfo {author} {\bibfnamefont {A.}~\bibnamefont
  {Mittelberger}},\ }\emph {\bibinfo {title} {Electron microscopy imaging of
  radiation-sensitive specimens with atomic resolution}},\ \href@noop {} {Ph.D.
  thesis},\ \bibinfo  {school} {University of Vienna} (\bibinfo {year}
  {2018})\BibitemShut {NoStop}%
\bibitem [{sup()}]{supplement}%
  \BibitemOpen
  \href@noop {} {}\bibinfo {note} {See Supplemental Material at HYPERLINK for
  additional figures regarding custom TEM chips and their rigidity, Raman
  spectroscopy pre-characterization, statistical distributions of the vacancy
  sizes and types, a comparison of the proposed model applied to experimental
  data with and without considering single vacancies and a visualization of the
  simulated graphene membranes containing no vacancies, single vacancies and
  double vacancies.}\BibitemShut {Stop}%
\bibitem [{\citenamefont {{Joudi}}\ \emph {et~al.}(2022)\citenamefont
  {{Joudi}}, \citenamefont {{Trentino}}, \citenamefont {{Mustonen}},
  \citenamefont {{Mangler}},\ and\ \citenamefont {{Kotakoski}}}]{Joudi2022}%
  \BibitemOpen
  \bibfield  {author} {\bibinfo {author} {\bibfnamefont {W.}~\bibnamefont
  {{Joudi}}}, \bibinfo {author} {\bibfnamefont {A.}~\bibnamefont {{Trentino}}},
  \bibinfo {author} {\bibfnamefont {K.}~\bibnamefont {{Mustonen}}}, \bibinfo
  {author} {\bibfnamefont {C.}~\bibnamefont {{Mangler}}}, \ and\ \bibinfo
  {author} {\bibfnamefont {J.}~\bibnamefont {{Kotakoski}}},\ }\href {\doibase
  10.1017/S1431927622009977} {\bibfield  {journal} {\bibinfo  {journal}
  {Microscopy and Microanalysis}\ }\textbf {\bibinfo {volume} {28 (S1)}},\
  \bibinfo {pages} {2626} (\bibinfo {year} {2022})}\BibitemShut {NoStop}%
\bibitem [{\citenamefont {Li}\ \emph {et~al.}(2016)\citenamefont {Li},
  \citenamefont {Kozbial}, \citenamefont {Nioradze}, \citenamefont {Parobek},
  \citenamefont {Shenoy}, \citenamefont {Salim}, \citenamefont {Amemiya},
  \citenamefont {Li},\ and\ \citenamefont {Liu}}]{Li2016}%
  \BibitemOpen
  \bibfield  {author} {\bibinfo {author} {\bibfnamefont {Z.}~\bibnamefont
  {Li}}, \bibinfo {author} {\bibfnamefont {A.}~\bibnamefont {Kozbial}},
  \bibinfo {author} {\bibfnamefont {N.}~\bibnamefont {Nioradze}}, \bibinfo
  {author} {\bibfnamefont {D.}~\bibnamefont {Parobek}}, \bibinfo {author}
  {\bibfnamefont {G.~J.}\ \bibnamefont {Shenoy}}, \bibinfo {author}
  {\bibfnamefont {M.}~\bibnamefont {Salim}}, \bibinfo {author} {\bibfnamefont
  {S.}~\bibnamefont {Amemiya}}, \bibinfo {author} {\bibfnamefont
  {L.}~\bibnamefont {Li}}, \ and\ \bibinfo {author} {\bibfnamefont
  {H.}~\bibnamefont {Liu}},\ }\href {\doibase 10.1021/acsnano.5b04843}
  {\bibfield  {journal} {\bibinfo  {journal} {ACS Nano}\ }\textbf {\bibinfo
  {volume} {10}},\ \bibinfo {pages} {349} (\bibinfo {year} {2016})}\BibitemShut
  {NoStop}%
\bibitem [{\citenamefont {Li}\ \emph {et~al.}(2013)\citenamefont {Li},
  \citenamefont {Wang}, \citenamefont {Kozbial}, \citenamefont {Shenoy},
  \citenamefont {Zhou}, \citenamefont {McGinley}, \citenamefont {Ireland},
  \citenamefont {Morganstein}, \citenamefont {Kunkel}, \citenamefont {Surwade},
  \citenamefont {Li},\ and\ \citenamefont {Liu}}]{Li2013}%
  \BibitemOpen
  \bibfield  {author} {\bibinfo {author} {\bibfnamefont {Z.}~\bibnamefont
  {Li}}, \bibinfo {author} {\bibfnamefont {Y.}~\bibnamefont {Wang}}, \bibinfo
  {author} {\bibfnamefont {A.}~\bibnamefont {Kozbial}}, \bibinfo {author}
  {\bibfnamefont {G.~J.}\ \bibnamefont {Shenoy}}, \bibinfo {author}
  {\bibfnamefont {F.}~\bibnamefont {Zhou}}, \bibinfo {author} {\bibfnamefont
  {R.}~\bibnamefont {McGinley}}, \bibinfo {author} {\bibfnamefont {P.~A.}\
  \bibnamefont {Ireland}}, \bibinfo {author} {\bibfnamefont {B.}~\bibnamefont
  {Morganstein}}, \bibinfo {author} {\bibfnamefont {A.}~\bibnamefont {Kunkel}},
  \bibinfo {author} {\bibfnamefont {S.~P.}\ \bibnamefont {Surwade}}, \bibinfo
  {author} {\bibfnamefont {L.}~\bibnamefont {Li}}, \ and\ \bibinfo {author}
  {\bibfnamefont {H.}~\bibnamefont {Liu}},\ }\href {\doibase 10.1038/nmat3709}
  {\bibfield  {journal} {\bibinfo  {journal} {Nature materials}\ }\textbf
  {\bibinfo {volume} {10}},\ \bibinfo {pages} {349} (\bibinfo {year}
  {2013})}\BibitemShut {NoStop}%
\bibitem [{\citenamefont {Amadei}\ \emph {et~al.}(2014)\citenamefont {Amadei},
  \citenamefont {Lai}, \citenamefont {Heskes},\ and\ \citenamefont
  {Chiesa}}]{Amadei2014}%
  \BibitemOpen
  \bibfield  {author} {\bibinfo {author} {\bibfnamefont {C.}~\bibnamefont
  {Amadei}}, \bibinfo {author} {\bibfnamefont {C.-Y.}\ \bibnamefont {Lai}},
  \bibinfo {author} {\bibfnamefont {D.}~\bibnamefont {Heskes}}, \ and\ \bibinfo
  {author} {\bibfnamefont {M.}~\bibnamefont {Chiesa}},\ }\href {\doibase
  10.1063/1.4893711} {\bibfield  {journal} {\bibinfo  {journal} {The Journal of
  chemical physics}\ }\textbf {\bibinfo {volume} {141}},\ \bibinfo {pages}
  {084709} (\bibinfo {year} {2014})}\BibitemShut {NoStop}%
\bibitem [{\citenamefont {Kozbial}\ \emph
  {et~al.}(2014{\natexlab{a}})\citenamefont {Kozbial}, \citenamefont {Li},
  \citenamefont {Sun}, \citenamefont {Gong}, \citenamefont {Zhou},
  \citenamefont {Wang}, \citenamefont {Xu}, \citenamefont {Liu},\ and\
  \citenamefont {Li}}]{Kozbial20141}%
  \BibitemOpen
  \bibfield  {author} {\bibinfo {author} {\bibfnamefont {A.}~\bibnamefont
  {Kozbial}}, \bibinfo {author} {\bibfnamefont {Z.}~\bibnamefont {Li}},
  \bibinfo {author} {\bibfnamefont {J.}~\bibnamefont {Sun}}, \bibinfo {author}
  {\bibfnamefont {X.}~\bibnamefont {Gong}}, \bibinfo {author} {\bibfnamefont
  {F.}~\bibnamefont {Zhou}}, \bibinfo {author} {\bibfnamefont {Y.}~\bibnamefont
  {Wang}}, \bibinfo {author} {\bibfnamefont {H.}~\bibnamefont {Xu}}, \bibinfo
  {author} {\bibfnamefont {H.}~\bibnamefont {Liu}}, \ and\ \bibinfo {author}
  {\bibfnamefont {L.}~\bibnamefont {Li}},\ }\href {\doibase
  10.1016/J.CARBON.2014.03.025} {\bibfield  {journal} {\bibinfo  {journal}
  {Carbon}\ }\textbf {\bibinfo {volume} {74}},\ \bibinfo {pages} {218}
  (\bibinfo {year} {2014}{\natexlab{a}})}\BibitemShut {NoStop}%
\bibitem [{\citenamefont {Kozbial}\ \emph
  {et~al.}(2014{\natexlab{b}})\citenamefont {Kozbial}, \citenamefont {Li},
  \citenamefont {Conaway}, \citenamefont {McGinley}, \citenamefont {Dhingra},
  \citenamefont {Vahdat}, \citenamefont {Zhou}, \citenamefont {Durso},
  \citenamefont {Liu},\ and\ \citenamefont {Li}}]{Kozbial20142}%
  \BibitemOpen
  \bibfield  {author} {\bibinfo {author} {\bibfnamefont {A.}~\bibnamefont
  {Kozbial}}, \bibinfo {author} {\bibfnamefont {Z.}~\bibnamefont {Li}},
  \bibinfo {author} {\bibfnamefont {C.}~\bibnamefont {Conaway}}, \bibinfo
  {author} {\bibfnamefont {R.}~\bibnamefont {McGinley}}, \bibinfo {author}
  {\bibfnamefont {S.}~\bibnamefont {Dhingra}}, \bibinfo {author} {\bibfnamefont
  {V.}~\bibnamefont {Vahdat}}, \bibinfo {author} {\bibfnamefont
  {F.}~\bibnamefont {Zhou}}, \bibinfo {author} {\bibfnamefont {B.}~\bibnamefont
  {Durso}}, \bibinfo {author} {\bibfnamefont {H.}~\bibnamefont {Liu}}, \ and\
  \bibinfo {author} {\bibfnamefont {L.}~\bibnamefont {Li}},\ }\href {\doibase
  10.1021/LA5018328} {\bibfield  {journal} {\bibinfo  {journal} {Langmuir}\
  }\textbf {\bibinfo {volume} {30}},\ \bibinfo {pages} {8598} (\bibinfo {year}
  {2014}{\natexlab{b}})}\BibitemShut {NoStop}%
\bibitem [{\citenamefont {Ashraf}\ \emph {et~al.}(2014)\citenamefont {Ashraf},
  \citenamefont {Wu}, \citenamefont {Wang}, \citenamefont {Aluru},
  \citenamefont {Dastgheib},\ and\ \citenamefont {Nam}}]{Ashraf2014}%
  \BibitemOpen
  \bibfield  {author} {\bibinfo {author} {\bibfnamefont {A.}~\bibnamefont
  {Ashraf}}, \bibinfo {author} {\bibfnamefont {Y.}~\bibnamefont {Wu}}, \bibinfo
  {author} {\bibfnamefont {M.~C.}\ \bibnamefont {Wang}}, \bibinfo {author}
  {\bibfnamefont {N.~R.}\ \bibnamefont {Aluru}}, \bibinfo {author}
  {\bibfnamefont {S.~A.}\ \bibnamefont {Dastgheib}}, \ and\ \bibinfo {author}
  {\bibfnamefont {S.}~\bibnamefont {Nam}},\ }\href {\doibase 10.1021/la503089k}
  {\bibfield  {journal} {\bibinfo  {journal} {Langmuir}\ }\textbf {\bibinfo
  {volume} {30}},\ \bibinfo {pages} {12827} (\bibinfo {year}
  {2014})}\BibitemShut {NoStop}%
\bibitem [{\citenamefont {Wei}\ and\ \citenamefont {Jia}(2015)}]{Wei2015}%
  \BibitemOpen
  \bibfield  {author} {\bibinfo {author} {\bibfnamefont {Y.}~\bibnamefont
  {Wei}}\ and\ \bibinfo {author} {\bibfnamefont {C.~Q.}\ \bibnamefont {Jia}},\
  }\href {\doibase 10.1016/J.CARBON.2015.02.019} {\bibfield  {journal}
  {\bibinfo  {journal} {Carbon}\ }\textbf {\bibinfo {volume} {87}},\ \bibinfo
  {pages} {10} (\bibinfo {year} {2015})}\BibitemShut {NoStop}%
\bibitem [{\citenamefont {M{\"u}cksch}\ \emph {et~al.}(2015)\citenamefont
  {M{\"u}cksch}, \citenamefont {R{\"o}sch}, \citenamefont {M{\"u}ller-Renno},
  \citenamefont {Ziegler},\ and\ \citenamefont {Urbassek}}]{Muksch2015}%
  \BibitemOpen
  \bibfield  {author} {\bibinfo {author} {\bibfnamefont {C.}~\bibnamefont
  {M{\"u}cksch}}, \bibinfo {author} {\bibfnamefont {C.}~\bibnamefont
  {R{\"o}sch}}, \bibinfo {author} {\bibfnamefont {C.}~\bibnamefont
  {M{\"u}ller-Renno}}, \bibinfo {author} {\bibfnamefont {C.}~\bibnamefont
  {Ziegler}}, \ and\ \bibinfo {author} {\bibfnamefont {H.~M.}\ \bibnamefont
  {Urbassek}},\ }\href {\doibase 10.1021/acs.jpcc.5b02948} {\bibfield
  {journal} {\bibinfo  {journal} {The Journal of Physical Chemistry C}\
  }\textbf {\bibinfo {volume} {119}},\ \bibinfo {pages} {12496} (\bibinfo
  {year} {2015})}\BibitemShut {NoStop}%
\bibitem [{\citenamefont {Smith}\ and\ \citenamefont
  {Lindley}(1998)}]{Smith1998}%
  \BibitemOpen
  \bibfield  {author} {\bibinfo {author} {\bibfnamefont {P.~J.}\ \bibnamefont
  {Smith}}\ and\ \bibinfo {author} {\bibfnamefont {P.~M.}\ \bibnamefont
  {Lindley}},\ }\href {\doibase 10.1063/1.56788} {\bibfield  {journal}
  {\bibinfo  {journal} {AIP Conference Proceedings}\ }\textbf {\bibinfo
  {volume} {449}},\ \bibinfo {pages} {133} (\bibinfo {year}
  {1998})}\BibitemShut {NoStop}%
\bibitem [{\citenamefont {Ponath}\ \emph {et~al.}(2013)\citenamefont {Ponath},
  \citenamefont {Posadas}, \citenamefont {Hatch},\ and\ \citenamefont
  {Demkov}}]{Ponath2013}%
  \BibitemOpen
  \bibfield  {author} {\bibinfo {author} {\bibfnamefont {P.}~\bibnamefont
  {Ponath}}, \bibinfo {author} {\bibfnamefont {A.}~\bibnamefont {Posadas}},
  \bibinfo {author} {\bibfnamefont {R.}~\bibnamefont {Hatch}}, \ and\ \bibinfo
  {author} {\bibfnamefont {A.}~\bibnamefont {Demkov}},\ }\href {\doibase
  10.1116/1.4798390} {\bibfield  {journal} {\bibinfo  {journal} {Journal of
  Vacuum Science Technology B: Microelectronics and Nanometer Structures}\
  }\textbf {\bibinfo {volume} {31}},\ \bibinfo {pages} {031201} (\bibinfo
  {year} {2013})}\BibitemShut {NoStop}%
\bibitem [{\citenamefont {Baker}(1980)}]{Baker1980}%
  \BibitemOpen
  \bibfield  {author} {\bibinfo {author} {\bibfnamefont {M.}~\bibnamefont
  {Baker}},\ }\href {\doibase https://doi.org/10.1016/0040-6090(80)90588-X}
  {\bibfield  {journal} {\bibinfo  {journal} {Thin Solid Films}\ }\textbf
  {\bibinfo {volume} {69}},\ \bibinfo {pages} {359} (\bibinfo {year}
  {1980})}\BibitemShut {NoStop}%
\bibitem [{\citenamefont {Chan}\ \emph {et~al.}(2001)\citenamefont {Chan},
  \citenamefont {Altman},\ and\ \citenamefont {Liang}}]{Chan2001}%
  \BibitemOpen
  \bibfield  {author} {\bibinfo {author} {\bibfnamefont {L.}~\bibnamefont
  {Chan}}, \bibinfo {author} {\bibfnamefont {E.}~\bibnamefont {Altman}}, \ and\
  \bibinfo {author} {\bibfnamefont {Y.}~\bibnamefont {Liang}},\ }\href
  {\doibase 10.1116/1.1367264} {\bibfield  {journal} {\bibinfo  {journal}
  {Journal of Vacuum Science Technology A}\ }\textbf {\bibinfo {volume} {19}},\
  \bibinfo {pages} {976} (\bibinfo {year} {2001})}\BibitemShut {NoStop}%
\bibitem [{\citenamefont {Oh}\ \emph {et~al.}(2016)\citenamefont {Oh},
  \citenamefont {Streller}, \citenamefont {Ashurst}, \citenamefont {Carpick},\
  and\ \citenamefont {Boer}}]{Oh2016}%
  \BibitemOpen
  \bibfield  {author} {\bibinfo {author} {\bibfnamefont {C.}~\bibnamefont
  {Oh}}, \bibinfo {author} {\bibfnamefont {F.}~\bibnamefont {Streller}},
  \bibinfo {author} {\bibfnamefont {W.~R.}\ \bibnamefont {Ashurst}}, \bibinfo
  {author} {\bibfnamefont {R.~W.}\ \bibnamefont {Carpick}}, \ and\ \bibinfo
  {author} {\bibfnamefont {M.~P.~D.}\ \bibnamefont {Boer}},\ }\href {\doibase
  10.1088/0960-1317/26/11/115020} {\bibfield  {journal} {\bibinfo  {journal}
  {Journal of Micromechanics and Microengineering}\ }\textbf {\bibinfo {volume}
  {26}},\ \bibinfo {pages} {115020} (\bibinfo {year} {2016})}\BibitemShut
  {NoStop}%
\bibitem [{\citenamefont {Isabell}\ \emph {et~al.}(1999)\citenamefont
  {Isabell}, \citenamefont {Fischione}, \citenamefont {O'Keefe}, \citenamefont
  {Guruz},\ and\ \citenamefont {Dravid}}]{Isabell1999}%
  \BibitemOpen
  \bibfield  {author} {\bibinfo {author} {\bibfnamefont {T.~C.}\ \bibnamefont
  {Isabell}}, \bibinfo {author} {\bibfnamefont {P.~E.}\ \bibnamefont
  {Fischione}}, \bibinfo {author} {\bibfnamefont {C.}~\bibnamefont {O'Keefe}},
  \bibinfo {author} {\bibfnamefont {M.~U.}\ \bibnamefont {Guruz}}, \ and\
  \bibinfo {author} {\bibfnamefont {V.~P.}\ \bibnamefont {Dravid}},\ }\href
  {\doibase 10.1017/S1431927699000094} {\bibfield  {journal} {\bibinfo
  {journal} {Microscopy and Microanalysis}\ }\textbf {\bibinfo {volume} {5}},\
  \bibinfo {pages} {126} (\bibinfo {year} {1999})}\BibitemShut {NoStop}%
\bibitem [{\citenamefont {Li}\ \emph {et~al.}(1997)\citenamefont {Li},
  \citenamefont {Belkind}, \citenamefont {Jansen},\ and\ \citenamefont
  {Orban}}]{Li1997}%
  \BibitemOpen
  \bibfield  {author} {\bibinfo {author} {\bibfnamefont {H.}~\bibnamefont
  {Li}}, \bibinfo {author} {\bibfnamefont {A.}~\bibnamefont {Belkind}},
  \bibinfo {author} {\bibfnamefont {F.}~\bibnamefont {Jansen}}, \ and\ \bibinfo
  {author} {\bibfnamefont {Z.}~\bibnamefont {Orban}},\ }\href {\doibase
  10.1016/S0257-8972(97)00079-0} {\bibfield  {journal} {\bibinfo  {journal}
  {Surface and Coatings Technology}\ }\textbf {\bibinfo {volume} {92}},\
  \bibinfo {pages} {171} (\bibinfo {year} {1997})}\BibitemShut {NoStop}%
\bibitem [{\citenamefont {Kotakoski}\ \emph {et~al.}(2014)\citenamefont
  {Kotakoski}, \citenamefont {Eder},\ and\ \citenamefont
  {Meyer}}]{Kotakoski2014}%
  \BibitemOpen
  \bibfield  {author} {\bibinfo {author} {\bibfnamefont {J.}~\bibnamefont
  {Kotakoski}}, \bibinfo {author} {\bibfnamefont {F.~R.}\ \bibnamefont {Eder}},
  \ and\ \bibinfo {author} {\bibfnamefont {J.~C.}\ \bibnamefont {Meyer}},\
  }\href {\doibase 10.1103/PhysRevB.89.201406} {\bibfield  {journal} {\bibinfo
  {journal} {Phys. Rev. B}\ }\textbf {\bibinfo {volume} {89}},\ \bibinfo
  {pages} {201406} (\bibinfo {year} {2014})}\BibitemShut {NoStop}%
\bibitem [{\citenamefont {Thiemann}\ \emph {et~al.}(2021)\citenamefont
  {Thiemann}, \citenamefont {Rowe}, \citenamefont {Zen}, \citenamefont
  {Müller},\ and\ \citenamefont {Michaelides}}]{Thiemann2021}%
  \BibitemOpen
  \bibfield  {author} {\bibinfo {author} {\bibfnamefont {F.~L.}\ \bibnamefont
  {Thiemann}}, \bibinfo {author} {\bibfnamefont {P.}~\bibnamefont {Rowe}},
  \bibinfo {author} {\bibfnamefont {A.}~\bibnamefont {Zen}}, \bibinfo {author}
  {\bibfnamefont {E.~A.}\ \bibnamefont {Müller}}, \ and\ \bibinfo {author}
  {\bibfnamefont {A.}~\bibnamefont {Michaelides}},\ }\href {\doibase
  10.1021/acs.nanolett.1c02585} {\bibfield  {journal} {\bibinfo  {journal}
  {Nano Letters}\ }\textbf {\bibinfo {volume} {21}},\ \bibinfo {pages} {8143}
  (\bibinfo {year} {2021})}\BibitemShut {NoStop}%
\bibitem [{\citenamefont {Lehtinen}\ \emph {et~al.}(2013)\citenamefont
  {Lehtinen}, \citenamefont {Kurasch}, \citenamefont {Krasheninnikov},\ and\
  \citenamefont {Kaiser}}]{Lehtinen2013}%
  \BibitemOpen
  \bibfield  {author} {\bibinfo {author} {\bibfnamefont {O.}~\bibnamefont
  {Lehtinen}}, \bibinfo {author} {\bibfnamefont {S.}~\bibnamefont {Kurasch}},
  \bibinfo {author} {\bibfnamefont {A.~V.}\ \bibnamefont {Krasheninnikov}}, \
  and\ \bibinfo {author} {\bibfnamefont {U.}~\bibnamefont {Kaiser}},\ }\href
  {\doibase 10.1038/ncomms3098} {\bibfield  {journal} {\bibinfo  {journal}
  {Nature Communications 2013 4:1}\ }\textbf {\bibinfo {volume} {4}},\ \bibinfo
  {pages} {1} (\bibinfo {year} {2013})}\BibitemShut {NoStop}%
\bibitem [{\citenamefont {Momma}\ and\ \citenamefont {Izumi}(2011)}]{VESTA}%
  \BibitemOpen
  \bibfield  {author} {\bibinfo {author} {\bibfnamefont {K.}~\bibnamefont
  {Momma}}\ and\ \bibinfo {author} {\bibfnamefont {F.}~\bibnamefont {Izumi}},\
  }\href {\doibase 10.1107/S0021889811038970} {\bibfield  {journal} {\bibinfo
  {journal} {Journal of Applied Crystallography}\ }\textbf {\bibinfo {volume}
  {44}},\ \bibinfo {pages} {1272} (\bibinfo {year} {2011})}\BibitemShut
  {NoStop}%
\bibitem [{\citenamefont {Babar}\ and\ \citenamefont
  {Kabir}(2018)}]{PhysRevB.98.075439}%
  \BibitemOpen
  \bibfield  {author} {\bibinfo {author} {\bibfnamefont {R.}~\bibnamefont
  {Babar}}\ and\ \bibinfo {author} {\bibfnamefont {M.}~\bibnamefont {Kabir}},\
  }\href {\doibase 10.1103/PhysRevB.98.075439} {\bibfield  {journal} {\bibinfo
  {journal} {Phys. Rev. B}\ }\textbf {\bibinfo {volume} {98}},\ \bibinfo
  {pages} {075439} (\bibinfo {year} {2018})}\BibitemShut {NoStop}%
\bibitem [{\citenamefont {Banhart}\ \emph {et~al.}(2011)\citenamefont
  {Banhart}, \citenamefont {Kotakoski},\ and\ \citenamefont
  {Krasheninnikov}}]{Banhart2011}%
  \BibitemOpen
  \bibfield  {author} {\bibinfo {author} {\bibfnamefont {F.}~\bibnamefont
  {Banhart}}, \bibinfo {author} {\bibfnamefont {J.}~\bibnamefont {Kotakoski}},
  \ and\ \bibinfo {author} {\bibfnamefont {A.~V.}\ \bibnamefont
  {Krasheninnikov}},\ }\href {\doibase 10.1021/nn102598m} {\bibfield  {journal}
  {\bibinfo  {journal} {ACS Nano}\ }\textbf {\bibinfo {volume} {5}},\ \bibinfo
  {pages} {26} (\bibinfo {year} {2011})}\BibitemShut {NoStop}%
\bibitem [{\citenamefont {Meyer}\ \emph {et~al.}(2007)\citenamefont {Meyer},
  \citenamefont {Geim}, \citenamefont {Katsnelson}, \citenamefont {Novoselov},
  \citenamefont {Booth},\ and\ \citenamefont {Roth}}]{Meyer2007}%
  \BibitemOpen
  \bibfield  {author} {\bibinfo {author} {\bibfnamefont {J.}~\bibnamefont
  {Meyer}}, \bibinfo {author} {\bibfnamefont {A.}~\bibnamefont {Geim}},
  \bibinfo {author} {\bibfnamefont {M.}~\bibnamefont {Katsnelson}}, \bibinfo
  {author} {\bibfnamefont {K.}~\bibnamefont {Novoselov}}, \bibinfo {author}
  {\bibfnamefont {T.}~\bibnamefont {Booth}}, \ and\ \bibinfo {author}
  {\bibfnamefont {S.}~\bibnamefont {Roth}},\ }\href {\doibase
  10.1038/nature05545} {\bibfield  {journal} {\bibinfo  {journal} {Nature}\
  }\textbf {\bibinfo {volume} {446}},\ \bibinfo {pages} {60} (\bibinfo {year}
  {2007})}\BibitemShut {NoStop}%
\bibitem [{\citenamefont {Fasolino}\ \emph {et~al.}(2007)\citenamefont
  {Fasolino}, \citenamefont {Los},\ and\ \citenamefont
  {Katsnelson}}]{Fasolino2007}%
  \BibitemOpen
  \bibfield  {author} {\bibinfo {author} {\bibfnamefont {A.}~\bibnamefont
  {Fasolino}}, \bibinfo {author} {\bibfnamefont {J.~H.}\ \bibnamefont {Los}}, \
  and\ \bibinfo {author} {\bibfnamefont {M.~I.}\ \bibnamefont {Katsnelson}},\
  }\href {\doibase 10.1038/nmat2011} {\bibfield  {journal} {\bibinfo  {journal}
  {Nature Materials}\ }\textbf {\bibinfo {volume} {6}},\ \bibinfo {pages} {858}
  (\bibinfo {year} {2007})}\BibitemShut {NoStop}%
\bibitem [{\citenamefont {Singh}\ \emph {et~al.}(2022)\citenamefont {Singh},
  \citenamefont {Scheinecker}, \citenamefont {Ludacka},\ and\ \citenamefont
  {Kotakoski}}]{Singh2022}%
  \BibitemOpen
  \bibfield  {author} {\bibinfo {author} {\bibfnamefont {R.}~\bibnamefont
  {Singh}}, \bibinfo {author} {\bibfnamefont {D.}~\bibnamefont {Scheinecker}},
  \bibinfo {author} {\bibfnamefont {U.}~\bibnamefont {Ludacka}}, \ and\
  \bibinfo {author} {\bibfnamefont {J.}~\bibnamefont {Kotakoski}},\ }\href
  {\doibase 10.3390/nano12203562} {\bibfield  {journal} {\bibinfo  {journal}
  {Nanomaterials}\ }\textbf {\bibinfo {volume} {12}},\ \bibinfo {pages} {3562}
  (\bibinfo {year} {2022})}\BibitemShut {NoStop}%
\bibitem [{\citenamefont {Joudi}\ \emph {et~al.}()\citenamefont {Joudi},
  \citenamefont {Windisch}, \citenamefont {Trentino}, \citenamefont {Propst},
  \citenamefont {Madsen}, \citenamefont {Susi}, \citenamefont {Mangler},
  \citenamefont {Mustonen}, \citenamefont {Libisch},\ and\ \citenamefont
  {Kotakoski}}]{phaidra}%
  \BibitemOpen
  \bibfield  {author} {\bibinfo {author} {\bibfnamefont {W.}~\bibnamefont
  {Joudi}}, \bibinfo {author} {\bibfnamefont {R.~S.}\ \bibnamefont {Windisch}},
  \bibinfo {author} {\bibfnamefont {A.}~\bibnamefont {Trentino}}, \bibinfo
  {author} {\bibfnamefont {D.}~\bibnamefont {Propst}}, \bibinfo {author}
  {\bibfnamefont {J.}~\bibnamefont {Madsen}}, \bibinfo {author} {\bibfnamefont
  {T.}~\bibnamefont {Susi}}, \bibinfo {author} {\bibfnamefont {C.}~\bibnamefont
  {Mangler}}, \bibinfo {author} {\bibfnamefont {K.}~\bibnamefont {Mustonen}},
  \bibinfo {author} {\bibfnamefont {F.}~\bibnamefont {Libisch}}, \ and\
  \bibinfo {author} {\bibfnamefont {J.}~\bibnamefont {Kotakoski}},\ }\href@noop
  {} {\enquote {\bibinfo {title} {Raw data for publication
  {C}orrugation-dominated mechanical softening of defect-engineered
  graphene},}\ }\bibinfo {note} {PHAIDRA (2025),
  https://phaidra.univie.ac.at/o:2119733}\BibitemShut {NoStop}%
\bibitem [{\citenamefont {Ferrari}\ \emph {et~al.}(2006)\citenamefont
  {Ferrari}, \citenamefont {Meyer}, \citenamefont {Scardaci}, \citenamefont
  {Casiraghi}, \citenamefont {Lazzeri}, \citenamefont {Mauri}, \citenamefont
  {Piscanec}, \citenamefont {Jiang}, \citenamefont {Novoselov}, \citenamefont
  {Roth},\ and\ \citenamefont {Geim}}]{Geim2006}%
  \BibitemOpen
  \bibfield  {author} {\bibinfo {author} {\bibfnamefont {A.}~\bibnamefont
  {Ferrari}}, \bibinfo {author} {\bibfnamefont {J.}~\bibnamefont {Meyer}},
  \bibinfo {author} {\bibfnamefont {V.}~\bibnamefont {Scardaci}}, \bibinfo
  {author} {\bibfnamefont {C.}~\bibnamefont {Casiraghi}}, \bibinfo {author}
  {\bibfnamefont {M.}~\bibnamefont {Lazzeri}}, \bibinfo {author} {\bibfnamefont
  {F.}~\bibnamefont {Mauri}}, \bibinfo {author} {\bibfnamefont
  {S.}~\bibnamefont {Piscanec}}, \bibinfo {author} {\bibfnamefont
  {D.}~\bibnamefont {Jiang}}, \bibinfo {author} {\bibfnamefont
  {K.}~\bibnamefont {Novoselov}}, \bibinfo {author} {\bibfnamefont
  {S.}~\bibnamefont {Roth}}, \ and\ \bibinfo {author} {\bibfnamefont
  {A.}~\bibnamefont {Geim}},\ }\href {\doibase 10.1103/PhysRevLett.97.187401}
  {\bibfield  {journal} {\bibinfo  {journal} {Physical review letters}\
  }\textbf {\bibinfo {volume} {97}},\ \bibinfo {pages} {187401} (\bibinfo
  {year} {2006})}\BibitemShut {NoStop}%
\bibitem [{\citenamefont {Tripathi}\ \emph {et~al.}(2017)\citenamefont
  {Tripathi}, \citenamefont {Mittelberger}, \citenamefont {Mustonen},
  \citenamefont {Mangler}, \citenamefont {Kotakoski}, \citenamefont {Meyer},\
  and\ \citenamefont {Susi}}]{Tripathi2017}%
  \BibitemOpen
  \bibfield  {author} {\bibinfo {author} {\bibfnamefont {M.}~\bibnamefont
  {Tripathi}}, \bibinfo {author} {\bibfnamefont {A.}~\bibnamefont
  {Mittelberger}}, \bibinfo {author} {\bibfnamefont {K.}~\bibnamefont
  {Mustonen}}, \bibinfo {author} {\bibfnamefont {C.}~\bibnamefont {Mangler}},
  \bibinfo {author} {\bibfnamefont {J.}~\bibnamefont {Kotakoski}}, \bibinfo
  {author} {\bibfnamefont {J.~C.}\ \bibnamefont {Meyer}}, \ and\ \bibinfo
  {author} {\bibfnamefont {T.}~\bibnamefont {Susi}},\ }\href {\doibase
  10.1002/pssr.201700124} {\bibfield  {journal} {\bibinfo  {journal} {physica
  status solidi ({RRL}) - Rapid Research Letters}\ }\textbf {\bibinfo {volume}
  {11}},\ \bibinfo {pages} {1700124} (\bibinfo {year} {2017})}\BibitemShut
  {NoStop}%
\bibitem [{\citenamefont {Dyck}\ \emph {et~al.}(2017)\citenamefont {Dyck},
  \citenamefont {Kim}, \citenamefont {Kalinin},\ and\ \citenamefont
  {Jesse}}]{Dyck2017}%
  \BibitemOpen
  \bibfield  {author} {\bibinfo {author} {\bibfnamefont {O.}~\bibnamefont
  {Dyck}}, \bibinfo {author} {\bibfnamefont {S.}~\bibnamefont {Kim}}, \bibinfo
  {author} {\bibfnamefont {S.~V.}\ \bibnamefont {Kalinin}}, \ and\ \bibinfo
  {author} {\bibfnamefont {S.}~\bibnamefont {Jesse}},\ }\href {\doibase
  10.1116/1.5003034} {\bibfield  {journal} {\bibinfo  {journal} {Journal of
  Vacuum Science Technology B}\ }\textbf {\bibinfo {volume} {36}},\ \bibinfo
  {pages} {011801} (\bibinfo {year} {2017})}\BibitemShut {NoStop}%
\bibitem [{\citenamefont {Egerton}\ \emph {et~al.}(2004)\citenamefont
  {Egerton}, \citenamefont {Li},\ and\ \citenamefont {Malac}}]{Egerton2004}%
  \BibitemOpen
  \bibfield  {author} {\bibinfo {author} {\bibfnamefont {R.}~\bibnamefont
  {Egerton}}, \bibinfo {author} {\bibfnamefont {P.}~\bibnamefont {Li}}, \ and\
  \bibinfo {author} {\bibfnamefont {M.}~\bibnamefont {Malac}},\ }\href
  {\doibase https://doi.org/10.1016/j.micron.2004.02.003} {\bibfield  {journal}
  {\bibinfo  {journal} {Micron}\ }\textbf {\bibinfo {volume} {35}},\ \bibinfo
  {pages} {399} (\bibinfo {year} {2004})}\BibitemShut {NoStop}%
\bibitem [{\citenamefont {McGilvery}\ \emph {et~al.}(2012)\citenamefont
  {McGilvery}, \citenamefont {Goode}, \citenamefont {Shaffer},\ and\
  \citenamefont {McComb}}]{Catriona2012}%
  \BibitemOpen
  \bibfield  {author} {\bibinfo {author} {\bibfnamefont {C.~M.}\ \bibnamefont
  {McGilvery}}, \bibinfo {author} {\bibfnamefont {A.~E.}\ \bibnamefont
  {Goode}}, \bibinfo {author} {\bibfnamefont {M.~S.}\ \bibnamefont {Shaffer}},
  \ and\ \bibinfo {author} {\bibfnamefont {D.~W.}\ \bibnamefont {McComb}},\
  }\href {\doibase https://doi.org/10.1016/j.micron.2011.10.026} {\bibfield
  {journal} {\bibinfo  {journal} {Micron}\ }\textbf {\bibinfo {volume} {43}},\
  \bibinfo {pages} {450} (\bibinfo {year} {2012})}\BibitemShut {NoStop}%
\bibitem [{\citenamefont {Dyck}\ \emph {et~al.}(2024)\citenamefont {Dyck},
  \citenamefont {Okmi}, \citenamefont {Xiao}, \citenamefont {Lei},
  \citenamefont {Lupini},\ and\ \citenamefont {Jesse}}]{Dyck2024}%
  \BibitemOpen
  \bibfield  {author} {\bibinfo {author} {\bibfnamefont {O.}~\bibnamefont
  {Dyck}}, \bibinfo {author} {\bibfnamefont {A.}~\bibnamefont {Okmi}}, \bibinfo
  {author} {\bibfnamefont {K.}~\bibnamefont {Xiao}}, \bibinfo {author}
  {\bibfnamefont {S.}~\bibnamefont {Lei}}, \bibinfo {author} {\bibfnamefont
  {A.~R.}\ \bibnamefont {Lupini}}, \ and\ \bibinfo {author} {\bibfnamefont
  {S.}~\bibnamefont {Jesse}},\ }\href {\doibase
  https://doi.org/10.1002/admi.202400598} {\bibfield  {journal} {\bibinfo
  {journal} {Advanced Materials Interfaces}\ ,\ \bibinfo {pages} {2400598}}
  (\bibinfo {year} {2024})}\BibitemShut {NoStop}%
\bibitem [{\citenamefont {Sader}\ \emph {et~al.}(1999)\citenamefont {Sader},
  \citenamefont {Chon},\ and\ \citenamefont {Mulvaney}}]{Sader1999}%
  \BibitemOpen
  \bibfield  {author} {\bibinfo {author} {\bibfnamefont {J.}~\bibnamefont
  {Sader}}, \bibinfo {author} {\bibfnamefont {J.}~\bibnamefont {Chon}}, \ and\
  \bibinfo {author} {\bibfnamefont {P.}~\bibnamefont {Mulvaney}},\ }\href
  {\doibase 10.1063/1.1150021} {\bibfield  {journal} {\bibinfo  {journal}
  {Review of Scientific Instruments}\ }\textbf {\bibinfo {volume} {70}},\
  \bibinfo {pages} {3967–3969} (\bibinfo {year} {1999})}\BibitemShut
  {NoStop}%
\bibitem [{\citenamefont {Perdew}\ \emph {et~al.}(1996)\citenamefont {Perdew},
  \citenamefont {Burke},\ and\ \citenamefont {Ernzerhof}}]{Perdew1996}%
  \BibitemOpen
  \bibfield  {author} {\bibinfo {author} {\bibfnamefont {J.~P.}\ \bibnamefont
  {Perdew}}, \bibinfo {author} {\bibfnamefont {K.}~\bibnamefont {Burke}}, \
  and\ \bibinfo {author} {\bibfnamefont {M.}~\bibnamefont {Ernzerhof}},\ }\href
  {\doibase 10.1103/PhysRevLett.77.3865} {\bibfield  {journal} {\bibinfo
  {journal} {Phys. Rev. Lett.}\ }\textbf {\bibinfo {volume} {77}},\ \bibinfo
  {pages} {3865} (\bibinfo {year} {1996})}\BibitemShut {NoStop}%
\bibitem [{\citenamefont {Kresse}\ and\ \citenamefont
  {Furthm\"uller}(1996)}]{VASP}%
  \BibitemOpen
  \bibfield  {author} {\bibinfo {author} {\bibfnamefont {G.}~\bibnamefont
  {Kresse}}\ and\ \bibinfo {author} {\bibfnamefont {J.}~\bibnamefont
  {Furthm\"uller}},\ }\href {\doibase 10.1103/PhysRevB.54.11169} {\bibfield
  {journal} {\bibinfo  {journal} {Phys. Rev. B}\ }\textbf {\bibinfo {volume}
  {54}},\ \bibinfo {pages} {11169} (\bibinfo {year} {1996})}\BibitemShut
  {NoStop}%
\end{thebibliography}%

\newpage\clearpage
\onecolumngrid
\setcounter{figure}{0}
\renewcommand{\figurename}{FIG.~S}

\section*{Supplemental Material}

\begin{figure}[h!]
\centering
\includegraphics[width=0.7\textwidth]{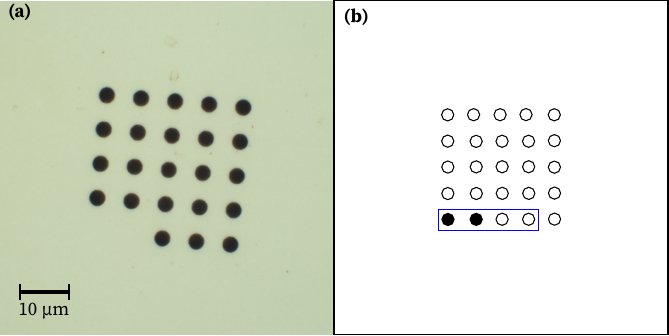}
\caption{\textbf{Finder grid.} (a) Visible light microscopy image of perforated SiN support membrane at 100$\times$ magnification and (b) sketch explaining the binary marking system (1100 corresponds to 3).}
\label{sfig:finder_grid}
\end{figure}

\begin{figure}[h!]
\centering
\includegraphics[width=\textwidth]{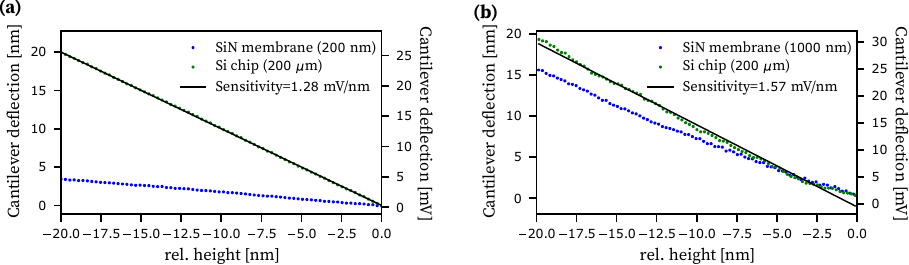}
\caption{\textbf{Support membrane rigidity.} (a) Comparison of force-distance curves acquired on Si frame ($200~\mu$m) and perforated SiN support membrane ($500 \times 500~\mu$m$^2$, $200$~nm thick) performed on commercially available PELCO holey SiN support film for TEM by Ted Pella, Inc. (b) Comparison of force-distance curves acquired on Si frame ($200~\mu$m) and perforated SiN support membrane ($90 \times 90~\mu$m$^2$, $1000$~nm thick) performed on custom SiN TEM support chip by Silson Ltd.}
\label{sfig:grid_bending}
\end{figure}

\begin{figure}[h!]
\centering
\includegraphics[width=\textwidth]{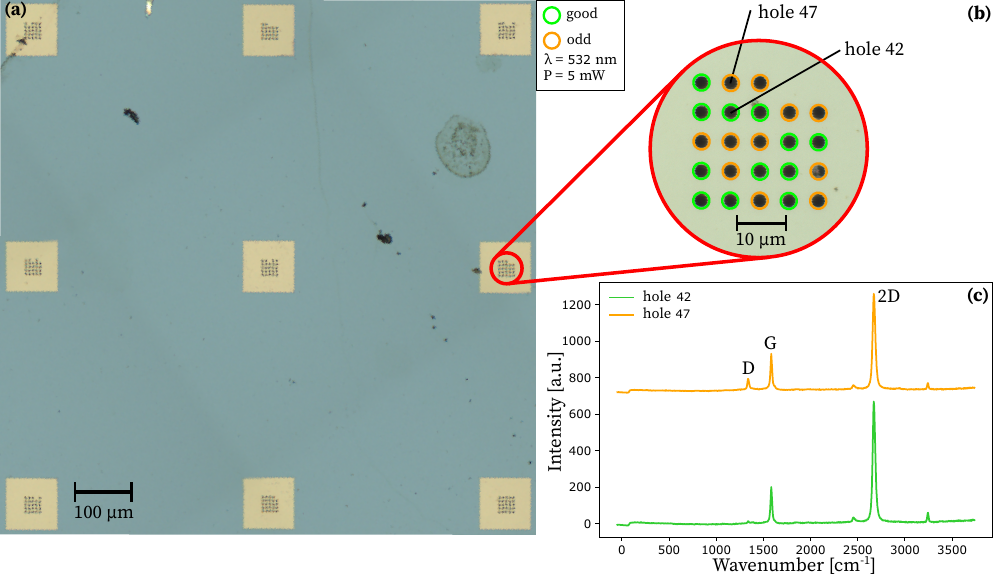}
\caption{\textbf{Raman pre-characterization.} (a) Visible light microscopy image of the sample, (b) magnified image of SiN support window with coloured overlay indicating the quality of the free-standing graphene above each hole and (c) corresponding example Raman spectra used for determining the local graphene quality.}
\label{sfig:Raman_prechar}
\end{figure}

\begin{figure}[h!]
\centering
\includegraphics[width=0.9\textwidth]{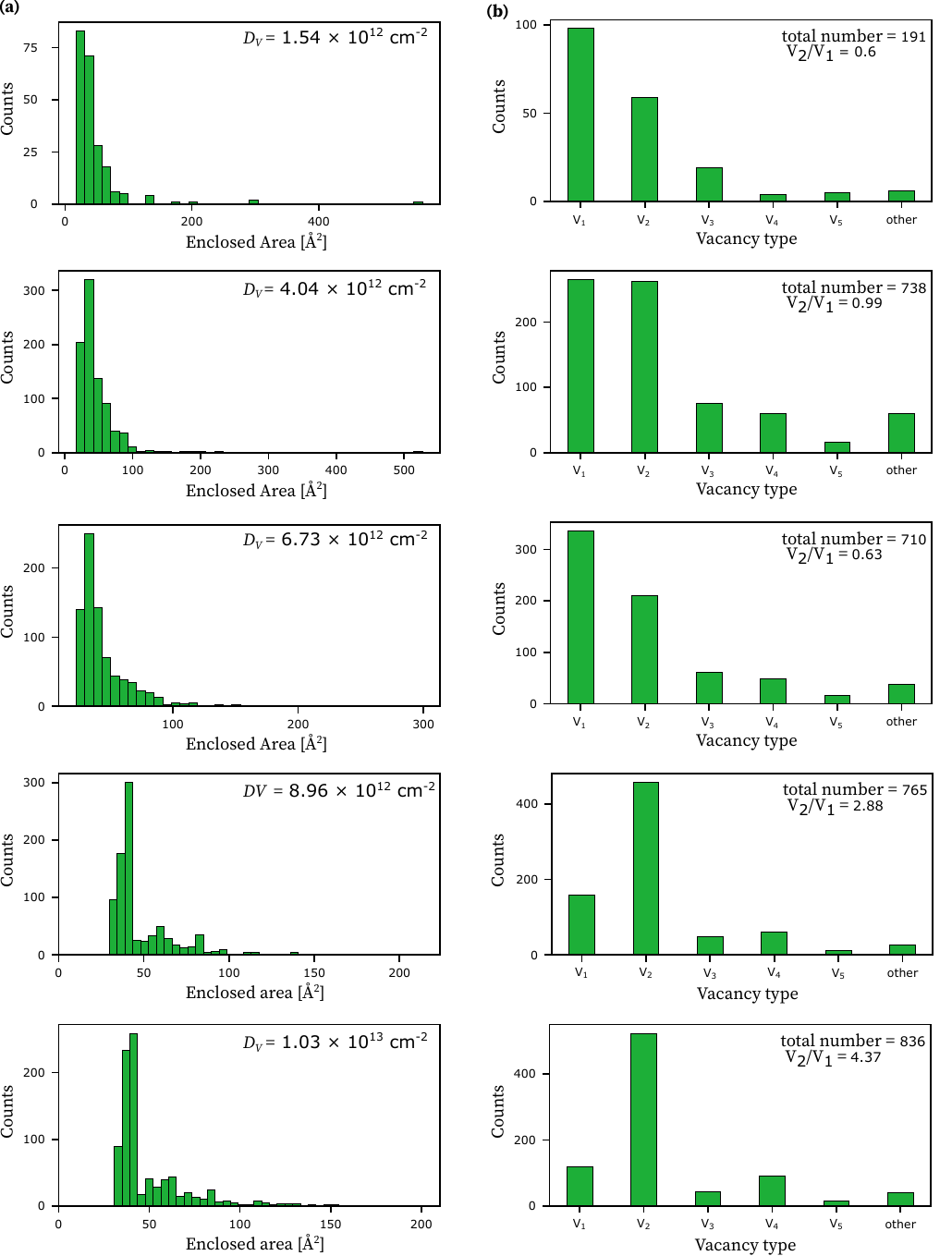}
\end{figure}

\begin{figure}[ht!]
\centering
\includegraphics[width=0.9\textwidth]{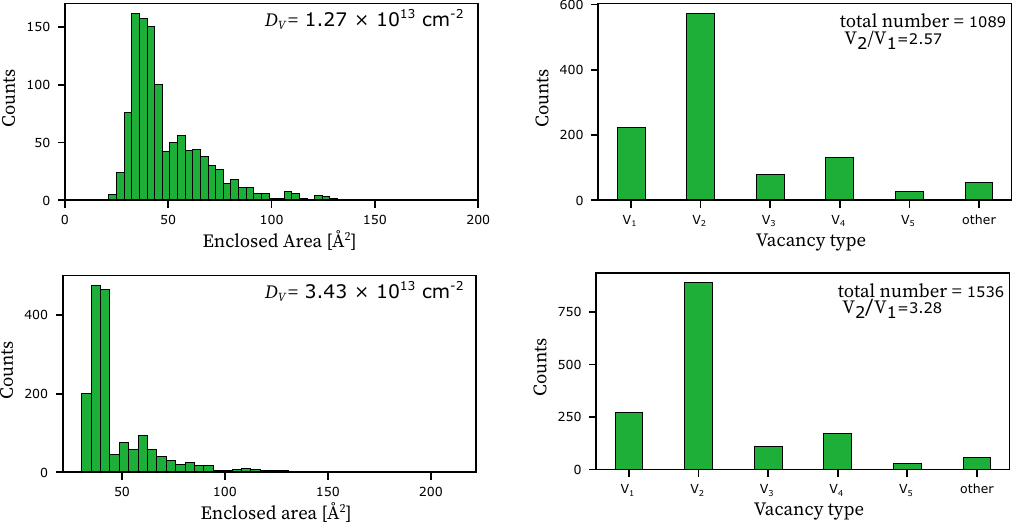}
\caption{\textbf{Vacancy distributions.} (a) Areal size distribution of the vacancies that were introduced into the the lattice, where the size is quantified by the area that is enclosed within the vacancy. $D_V$ represents the vacancy density of the corresponding sample. (b) Vacancy-type distribution, where the subscript represents the number of missing carbon atoms in the respective vacancy structure.}
\label{sfig:CNN_statistics_summary}
\end{figure}

\begin{figure}[hb!]
\centering
\includegraphics[width=.7\textwidth]{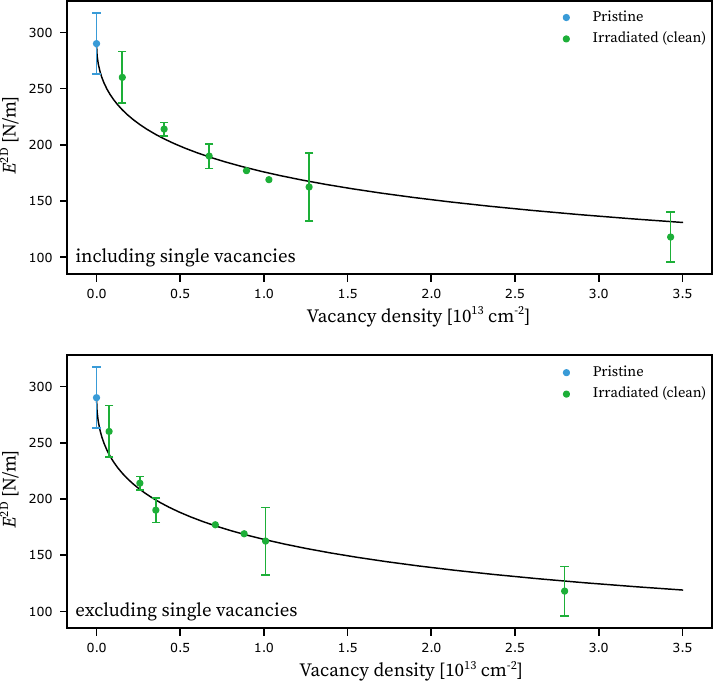}
\caption{\textbf{Vacancy type effect.} Model accounting for corrugation compared to data where single vacancies have been included in (top) and excluded from (bottom) the total vacancy density.}
\label{sfig:single_vs_double}
\end{figure}

\begin{figure}
    \centering
    \includegraphics[width=1\textwidth]{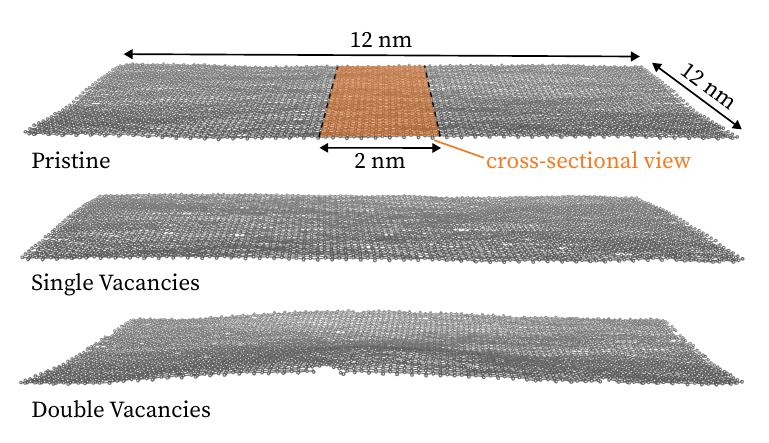}
    \caption{\textbf{Simulated graphene membranes.} Visualization of the simulated structures of pristine graphene, graphene with single vacancies and graphene with double vacancies, each containing up to $6,000$ atoms. The pristine case includes a schematic illustration of the viewing region shown in Fig.~\ref{fig: corrugation_sim}(b). The visualizations have been created by the VESTA software~\cite{VESTA}.}
    \label{sfig:corrugation_visualized}
\end{figure}

\end{document}